%% file: main.tex
\documentclass[leqno,12pt,a4paper]{article}
\usepackage{setspace}
\usepackage{latexsym}
\usepackage{amsmath}
\usepackage{amssymb}
\usepackage{mathtools}
\usepackage{bbm}
\usepackage{amsthm}
\usepackage{marvosym}

\usepackage[]{algorithm,algpseudocode}
\algnewcommand\algorithmicforeach{\textbf{for each}}
\algdef{S}[FOR]{ForEach}[1]{\algorithmicforeach\ #1\ \algorithmicdo}

\usepackage{graphicx}
\usepackage{epsfig}
\usepackage{subcaption}
\usepackage{tabularx}
\newcolumntype{L}[1]{>{\raggedright\arraybackslash}p{#1}} 
\newcolumntype{C}[1]{>{\centering\arraybackslash}p{#1}} 
\newcolumntype{R}[1]{>{\raggedleft\arraybackslash}p{#1}} 

\usepackage{array}

\allowdisplaybreaks
\usepackage{tikz}
\usetikzlibrary{shapes,arrows,patterns,backgrounds,fit,arrows.meta}
\usepackage{pgfplots}
\pgfplotsset{
  compat=newest,
  width=0.65\textwidth,
  height=0.4\textwidth,
  tick align=outside,
  tick pos=left,
  scaled ticks=false,
  grid=major,
  grid style={dotted},
}
\pgfkeys{/pgf/number format/.cd,
  1000 sep={\,}, 
  min exponent for 1000 sep=4,
}

\usepackage{siunitx}
\usepackage{makecell}

\usepackage{comment}

\newcommand{\boldface}[1]{\boldsymbol{#1}}

\newcommand{\bfb}{\boldface{b}}

\newcommand{\bfu}{\boldface{u}}

\newcommand{\bfx}{\boldface{x}}
\newcommand{\bfy}{\boldface{y}}

\newcommand{\bfB}{\boldface{B}}

\newcommand{\bfF}{\boldface{F}}

\newcommand{\bfI}{\boldface{I}}

\newcommand{\bfN}{\boldface{N}}

\newcommand{\bfvarepsilon}{\boldsymbol{\varepsilon}}

\newcommand{\bfsigma}{\boldsymbol{\sigma}}

\usepackage{bbold}

\newcommand{\dsE}{\mathbb{E}}

\newcommand{\be}{\begin{equation}}
\newcommand{\ee}{\end{equation}}
\newcommand{\bea}{\begin{eqnarray}}
\newcommand{\eea}{\end{eqnarray}}
\newcommand{\bes}{\begin{equation*}}
\newcommand{\ees}{\end{equation*}}
\newcommand{\beas}{\begin{eqnarray*}}
\newcommand{\eeas}{\end{eqnarray*}}

\newcommand{\dd}{\ \mathrm{d}}

\usepackage{lipsum}
\usepackage{placeins}

\definecolor{RUBblue}{rgb}{0.0470588,0.262745,0.411765}
\definecolor{lightgray}{gray}{0.92}
\definecolor{blau}{rgb}{0,0.25,1.0}
\definecolor{RUBCDblue}{rgb}{0.062745,0.305882,0.545098}
\definecolor{RUBCDgreen}{rgb}{0.517647,0.741176,0.0}
\definecolor{RUBCDgray}{rgb}{0.815686,0.815686,0.807843}
\definecolor{RUBCDgrayDark}{rgb}{0.615686,0.615686,0.607843}

\makeatletter

\def\env@cases#1{%
  \let\@ifnextchar\new@ifnextchar
  \left\lbrace\def\arraystretch{1.2}%
  \array{@{}#1@{\quad}l@{}}}
\makeatother

\usepackage{booktabs}

\RequirePackage{relsize}
\newfont{\Sf}{cmssbx10 scaled 2074}

\usepackage[
  colorinlistoftodos,
  backgroundcolor=green,
  linecolor=green,
  textsize=footnotesize
]{todonotes}

\usepackage{import}
\graphicspath{{Figures/}}

\newcommand{\centered}[1]{\begin{tabular}{l} #1 \end{tabular}}

\newcommand{\rev}[1]{\textcolor{black}{#1}}

\usepackage[backend=bibtex,firstinits=true, url=false]{biblatex}
\begin{document}
\title{Efficient damage simulations under material uncertainties in a weakly-intrusive implementation}
\date{} 
\maketitle
{\large
\noindent{Hendrik}  Geisler$^{1, 2, 3}$, Emmanuel Baranger$^{2,3}$, Philipp Junker$^{1, 2}$\\[0.5mm]
}
1: Leibniz University Hannover, Institute of Continuum Mechanics, Hannover, Germany\\
2: IRTG 2657: Computational Mechanics Techniques in High Dimensions \\
3: Université Paris-Saclay, CentraleSupélec, ENS Paris-Saclay, CNRS, LMPS - Laboratoire de Mécanique Paris-Saclay, Gif-sur-Yvette, France \\[2mm]
\Letter \hspace{0.1cm} geisler@ikm.uni-hannover.de

\section*{Abstract}
Uncertainty quantification is not yet widely adapted in the design process of engineering components despite its importance for achieving sustainable and resource-efficient structures. This is mainly due to two reasons:
1) Tracing the effect of uncertainty in engineering simulations is a computationally challenging task. This is especially true for inelastic simulations as the whole loading history influences the results.
2) Implementations of efficient schemes in standard finite element software are lacking. \\
In this paper, we are tackling both problems. We are proposing a \rev{weakly}-intrusive implementation of the time-separated stochastic mechanics in the finite element software Abaqus. The time-separated stochastic mechanics is an efficient and accurate method for the uncertainty quantification of structures with inelastic material behavior. The method effectivly separates the stochastic but time-independent from the deterministic but time-dependent behavior.
The resulting scheme consists only two deterministic finite element simulations for homogeneous material fluctuations in order to approximate the stochastic behavior. This brings down the computational cost compared to standard Monte Carlo simulations by at least two orders of magnitude while ensuring accurate solutions.
In this paper, the implementation details in Abaqus and numerical comparisons are presented for the example of damage simulations.
\newline
Keywords: Time-separated stochastic mechanics, \rev{weakly}-intrusive, damage, Abaqus

\section{Introduction}
Fluctuations in the material parameters have a significant influence on the response of structures and consequently their reliability. Therefore, the consideration of uncertainty in the simulation of structures is of ever increasing interest. 
Unfortunately, suitable computationally efficient methods and their readily available implementations are still lacking. This is especially true for inelastic material behavior as damage.

For linear elastic material behavior methods for uncertainty quantification are known in the framework of Stochastic Finite Element methods \cite{stefanou_stochastic_2008, sudret_stochastic_2000}. These are based on sampling methods as Monte Carlo, and methods to increase its computational efficiency as Stochastic Collocation, Perturbation and Spectral Stochastic FEM.
The underlying idea of these methods is to approximate the response surface by a suitable basis in order to accelerate the model evaluation.
A particularly mature environment to test these methods is UQLab \cite{marelli_uqlab_2014}.

\begin{figure}
	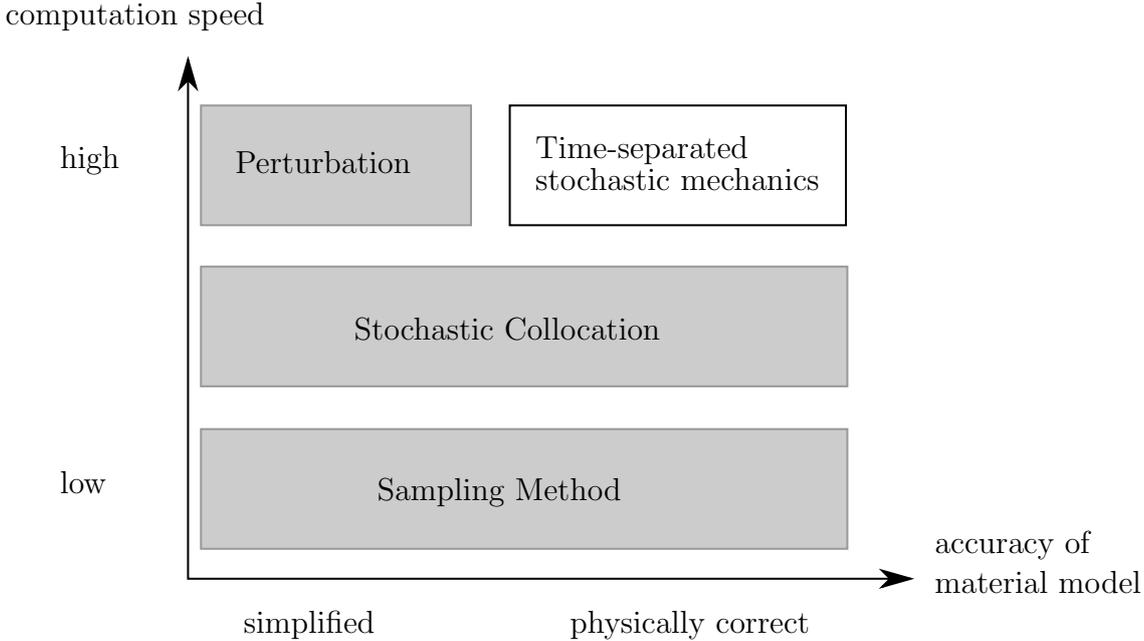
	\caption{Literature overview of methods for uncertainty quantification of structures with inelastic material behavior. Grey boxes represent existing techniques.}
	\label{fig:LiteratureOverview}
\end{figure}

For uncertainty quantification of structures with inelastic material behavior, mainly thee different techniques have been used.
These techniques are explained in more detail below.
\begin{description}
    \item[Sampling method] Most sampling methods are variants of the Monte Carlo scheme \cite{liu_monte_2004, ghanem_hybrid_1998}. The methods are based on a repeated evaluation of the system each time with a different realization of the random variables.
    Unfortunately, the results are only converged for a high enough number of iterations rendering the method rather computationally expensive.
    \item[Stochastic Collocation] A more intricate choice of the sampling points, e.g., Smolyak points \cite{xiu_high-order_2005} allows to reduce the number of needed individual simulations. For interpolation between the sampling points, a reduced-order polynomial surface is used \cite{xiu_stochastic_2016, ghanem_sparse_2017}. Nevertheless, the method suffers from reduced efficiency for high-dimensional problems and instability \cite{dannert_investigations_2022, feng_performance_2021}.
    \item[Perturbation method] The response surface can also be build by means of a Taylor series expansion around the expectation of the random variables \cite{kleiber_stochastic_1992, kaminski_generalized_2010}. This allows the design of a very efficient method. Unfortunately, only simplified polynomial material models can be handled.
\end{description}
The \rev{discussed} methods are presented in Figure~\ref{fig:LiteratureOverview}. The methods are ordered by computational speed and the complexity of the possible material models to be investigated. It is obvious, that a method combining high computational speed with high accuracy is still missing. The time-separated stochastic mechanics was developed to close this gap in literature.

Likewise, in the field of damage mechanics, uncertainty quantification is a reoccurring theme. Thus, many authors have tried the above mentioned method and provided their analysis. References for the use of (Markov Chain) Monte Carlo methods are found in \cite{troy_uncertainty_2023} and \cite{barros_de_moraes_integrated_2021}. The stochastic collocation method found repeated use in, e.g., \cite{liu_monte_2004} and \cite{dos_santos_oliveira_sparse_2024}. 

The main complexity in dealing with inelastic material behavior comes from the fact that additional internal variables with accompanying evolution equation are introduced. These internal variables capture the microstructural state. Therefore, a system of coupled (partial) differential equations have to be solved. In addition, the whole loading history influences the current values of all fields, i.e. strain, internal variables and stress.
Therefore, the design of a suitable approximation scheme for uncertainty quantification is not trivial. 
The time-separated stochastic mechanics (TSM) (\cite{geisler_new_2024, junker_modeling_2020}) reduces the computational complexity by separating the time-dependent but deterministic from the stochastic but time-independent behavior of each field. The resulting scheme only needs a low number of finite element simulations, depending on the problem at hand. The uncertainty quantification can be efficiently performed in a post-processing step.

The method has shown astonishing success for the dynamic simulation of a visco-elastic material~\cite{geisler_simulation_2022} under material uncertainties and for visco-elastic structures with local material fluctuations \cite{geisler_time-separated_2023} and for the homogenization of composites \cite{geisler_uncertainty_2024}. Recently, the TSM was investigated for a large class of material models on the material point level \cite{geisler_new_2024}.
Nevertheless, the method \rev{has only been} implemented in an in-house code \rev{so far}. Thus, the efficient implementation of the TSM in a commercial finite element program is \rev{presented}.

\subsection{Contribution}
\rev{As an example for a weakly-intrusive } implementation of the time-separated stochastic mechanics in the finite element software Abaqus\rev{, a} viscous damaging material model with homogenous material fluctuations is chosene.
In total, only two deterministic finite element simulations combined with a fast post-processing need to be performed. Nevertheless, the results are remarkably similar to a reference Monte Carlo solution with hundreds of iterations \rev{necessary for convergence}.
The presented scheme is \rev{weakly}-intrusive and makes use of various user-defined subroutines in Abaqus and the Python scripting interface. 

\subsection{Structure}
This work is structured as follows: \rev{t}he material model \rev{for damage} and the modeling assumptions are presented in Section~\ref{sec:MatModel}. The time-separated stochastic mechanics is explained in Section~\ref{sec:TSM}. The implementation in Abaqus is discussed in Section~\ref{sec:ImplAbaqSec}. Numerical results are presented in Section~\ref{sec:NumericalResults}. The computational efficiency is analyzed in Section~\ref{sec:CompEff}. The conclusion\rev{s are} given in Section~\ref{sec:Conclusion}.

\section{Material model and modeling assumptions}
\label{sec:MatModel}
In literature, various different damage models have been proposed\rev{, e.g., \cite{besson_continuum_2010,pijaudier-cabot_review_2014,de_borst_fracture_2002}}. A typical damage model is characterized by two factors: loss of stiffness for a high loading and mesh-independence. Individual damage models differ in the driving force, e.g., energy or stress, and the strategy to circumvent mesh-dependency, e.g., by gradient enhancement or by rate-limitation.
For this study, we choose a simple model which possesses the above mentioned characteristics: a viscous-type damage model.
The damage state is captured by a scalar-valued internal variable $d$ with $d=0$ indicating the undamaged state.
A viscous evolution of the damage variable is given by
\begin{equation}
    \dot{d}(t) = \frac{1}{\eta} f(d) \Psi_0(t) \label{eq:damagediffeq}
\end{equation}
with the viscosity $\eta$ and the damage function $f(d) = \exp(-d(t)) \in (0, 1]$. A value of $f = 1$ represents an undamaged material, whereas $f \to 0$ indicates a fully damaged material.
The Helmholtz free energy $\Psi_0$ of the undamaged state is given as
\begin{equation}
    \Psi_0 = \frac{1}{2} \bfvarepsilon \cdot \dsE \cdot \bfvarepsilon
\end{equation}
with the strains $\bfvarepsilon$ and the elasticity tensor $\dsE$.
The usual balance of linear momentum in dynamics is given by
\begin{equation}
    \nabla \cdot \bfsigma + \bfb^* = \rho \ddot{\bfu}
\end{equation} 
with prescribed (body) forces $\bfb^*$, acceleration $\ddot{\bfu}$, density $\rho$ and the stress
\begin{equation}
    \bfsigma = f(d) \, \dsE \cdot \bfvarepsilon. \label{eq:stresseq}
\end{equation}

The damage evolution $\dot{d}(t)$ in Equation~\eqref{eq:damagediffeq} is positive, reflecting that no healing of the material can occur.
The mesh-independence of the material model is achieved due to the viscous evolution of the damage variable. An internal length is introduced in dynamic simulations \cite{bazant_nonlocal_2002, needleman_material_1988} such that nonlocality results. The internal length is given by
\begin{equation}
    \ell_v = \frac{\eta}{v \rho} = \frac{\eta}{\sqrt{E \rho}}
\end{equation}
where $v = \sqrt{E / \rho}$ is the longitudinal wave propagation velocity. The Young's modulus is given as $E$ and the density as $\rho$. Equivalently, the characteristic time $\tau_0$, which a wave front takes to pass over the distance $\ell_v$, can be stated as
\begin{equation}
    \tau_0 = \frac{\eta}{E}.
\end{equation}

The above equations are sufficient to describe the response of a structure given suitable boundary conditions and a set of deterministic material parameters $m = \{\eta, E, \nu, \rho\}$.
However, for real-life structures the material parameters are not fixed but are fluctuating between specimens.
This is due to a variety of reason as variations in the production process or material irregularities.
The fluctuations of the material parameter are a main source of uncertainty and complicate the development of sustainable and resource-efficient designs.
In this work, we are assuming random homogeneous fluctuations of Young's modulus and Poisson's ratio which lead to a random elasticity tensor and consequently uncertain displacement, stress and reaction force. These fluctuations can be modeled by a dependency of the elasticity tensor $\dsE$ on a random variable $\xi$. To simplify the mathematical description the random variable $\xi$ is zero-centered, i.e., $\langle \xi \rangle = 0$ with the expectation operator $\langle \cdot \rangle$. We denote the expectation of the elasticity tensor by $\dsE^{(0)} := \langle \dsE \rangle$. With these preliminaries we can express the random elasticity tensor as
\begin{equation}
	\dsE = (1 + \xi) \, \dsE^{(0)}. \label{eq:Erand}
\end{equation}
In the following, the standard deviation of a quantity is denoted as $\textrm{Std}(\cdot)$.

\section{Time-separated stochastic mechanics}
\label{sec:TSM}
The time-separated stochastic mechanics (TSM) \cite{junker_modeling_2020, geisler_new_2024} is an efficient and accurate method for the uncertainty quantification of structures with inelastic material models. The uncertain parameters, e.g., material parameters as the Young's modulus, render all quantities as strains, stresses and internal variables random. This complicates the solution procedure compared to the deterministic case enormously.

In deterministic finite element simulations, the balance of linear momentum and evolution equation only have to be solved once to calculate the value of all quantities over time.
In the presence of uncertainties, all equations and fields are dependent on one or multiple random variables $\xi_j$. Thus, the uncertainty quantification of the fields, e.g., calculation of expectation and standard deviation, requires the solution of the equation system for a large number of realizations of $\xi_j$. This quickly gets computationally intractable.

The TSM reduces the computational complexity by the use of a separated representation for all field variables. Any field $\bfy = f(t, \bfx, \xi)$ is approximated by a separated expansion in the form
\begin{equation}
	\bfy(t, \bfx, \xi) = \bfy^{(0)}(t, \bfx) + \sum_{i=1}^p \bfI^{(i)}(t, \bfx) \, \xi^i \label{eq:TSMgeneral}
\end{equation}
with 
\begin{equation}
	\bfy^{(0)}(t, \bfx) = \bfy(t, \bfx, \xi) \big\vert_{\xi=0}
\end{equation}
for the case of a scalar zero-centered random variable $\xi$. The general case is presented in \cite{geisler_new_2024} \rev{for investigations on the material point}. The expansion in Equation~\eqref{eq:TSMgeneral} effectively separates the stochastic but time-independent behavior ($\xi^i$) from the deterministic but time-dependent behavior ($\bfy^0(t), \bfI^{(1)}(t), I^{(2)}(t) \dots$). This allows to only calculate the deterministic but time-dependent behavior via FEM simulations and account for the stochastic behavior during post-processing. It may be remarked that the order $p$ of the approximation is typically very low, i.e., p = \{1,2\} such that only a low number of FEM simulations is necessary.
Depending on the governing equation of the field $\bfy$, the deterministic but time-dependent terms $\bfI^{(i)}$ can be identified. For an algebraic equation as the balance of linear momentum, the terms $\bfI^{(i)}$ are simply given as Taylor series term, i.e.,
\begin{equation}
	\bfI^{(i)} = \frac{\dd^i y}{\dd \xi^i} \Big\vert_{\xi = 0}.
\end{equation} 
For the case \rev{when} $\bfy(t)$ is given \rev{implicitly by a solution of a } differential equation, e.g., as for the evolution equation of an internal variable, the determination of $\bfI^{(i)}$ is more intricate. The full procedure is presented in~\cite{geisler_new_2024} \rev{for investigations on the material point}.

The advantage of the above scheme is obvious. The separation of time-dependent and stochastic behavior allows to perform a low number of deterministic simulation to determine the terms $\bfy^0(t, \bfx)$ and $\bfI^{(i)}(t, \bfx)$ and estimate the effect of the random variable $\xi$ only afterwards. For homogeneous material fluctuations and a linear Taylor series, i.e., $p = 1$, only one additional finite element simulation is needed compared to the fully deterministic case.

\begin{figure}
	\includegraphics[trim=1.6cm 17cm 0cm 2.0cm, clip, scale=1.]{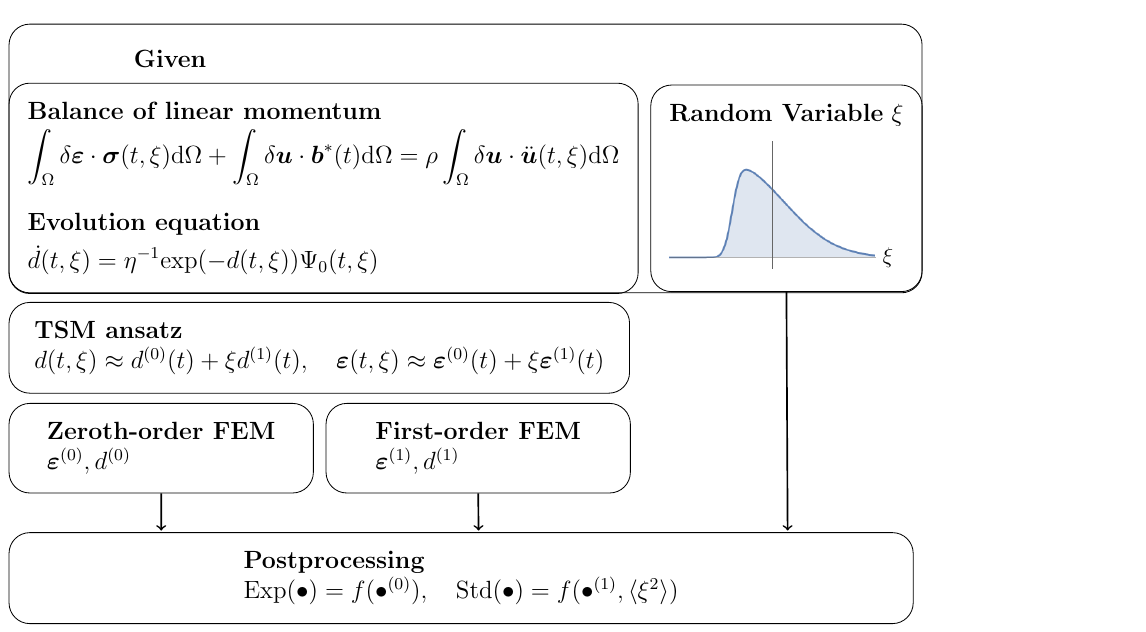}
	\caption{Graphical overview of the time-separated stochastic mechanics approach.}
	\label{fig:GraphAbstract}
\end{figure}

The overall procedure is summarized in Figure~\ref{fig:GraphAbstract}. Here, the TSM is already applied to the problem at hand: viscous damage simulations. 

\subsection{Separated representation}
\label{sec:SeparatedRepresentation}
A linear series expansion for the displacement field $\bfu(\bfx, t, \xi)$ in the random variable $\xi$ reads as
\begin{equation}
    \bfu^\textrm{TSM}(\bfx, t, \xi) = \bfu^{(0)}(\bfx, t) + \xi \bfu^{(1)}(\bfx, t) \approx \bfu(\bfx, t, \xi). \label{eq:uTSM}
\end{equation}

Of course, the displacement field $\bfu$ depends on the coordinate $\bfx$, the time $t$ and additionally the random variable $\xi$. The series expansion in Equation~\eqref{eq:uTSM} is a linear approximation of the displacement with respect to the random variable $\xi$. The displacement field for $\xi = 0$ is given by $\bfu^{(0)}(\bfx, t)$. The (linear) change of the displacement field with respect to the random variable is given by $\bfu^{(1)}(\bfx, t)$. The fields $\bfu^{(0)}(\bfx, t)$ and $\bfu^{(1)}(\bfx, t)$ have to be determined.

The strain field is given by the usual relationship $\bfvarepsilon = \nabla \bfu$ and reads as
\begin{equation}
    \bfvarepsilon^\textrm{TSM}(\bfx, t, \xi) = \bfvarepsilon^{(0)}(\bfx, t) + \xi \bfvarepsilon^{(1)}(\bfx, t) \approx \bfvarepsilon(\bfx, t, \xi)
\end{equation}
where the upper index $\textrm{TSM}$ indicates the approximation.

As we are dealing with a dynamic problem, similar series expansions are used for the velocity and acceleration fields. As the random variable is not time-dependent the series expansion for velocity and acceleration result as the time-derivative
\begin{equation}
\dot{\bfu}^\textrm{TSM}(\bfx, t, \xi) = \dot{\bfu}^{(0)}(\bfx, t) + \xi \dot{\bfu}^{(1)}(\bfx, t) \approx \dot{\bfu}(\bfx, t, \xi)
\end{equation}
and similarly
\begin{equation}
\ddot{\bfu}^\textrm{TSM}(\bfx, t, \xi) = \ddot{\bfu}^{(0)}(\bfx, t) + \xi \ddot{\bfu}^{(1)}(\bfx, t) \approx \ddot{\bfu}(\bfx, t, \xi).
\end{equation}

The internal damage variable is also random. Consequently, we use the same type of expansion to represent the internal damage variable as
\begin{equation}
    d^\textrm{TSM}(\bfx, t, \xi) = d^{(0)}(\bfx, t) + \xi d^{(1)}(\bfx, t) \approx d(\bfx, t, \xi). \label{eq:dTSM}
\end{equation}

With these approximations the stress is given as
\begin{equation}
	\bfsigma = f(d) \, \dsE \cdot \bfvarepsilon \approx f(d^\textrm{TSM}) (1 + \xi) \, \dsE^{(0)} \cdot \left(\bfvarepsilon^{(0)}(\bfx, t) + \xi \bfvarepsilon^{(1)}(\bfx, t) \right).
\end{equation}
Obviously, the stress is highly nonlinear in the random variable $\xi$.

\subsubsection{Identification of terms: balance equation}
\label{sec:BalanceEquation}
It remains to define the equations to calculate the strain fields $\bfvarepsilon^{(0)}(t, \bfx)$ and $\bfvarepsilon^{(1)}(t, \bfx)$.
For this, we employ the balance of linear momentum as for the deterministic case. The balance equation in weak form with the approximations above reads as
\begin{equation}
    \int_\Omega \delta\bfvarepsilon^{(0)} \cdot \bfsigma(\xi) \dd \Omega + \int_\Omega \delta\bfu^{(0)} \cdot \bfb^* \dd\Omega = \rho \int_\Omega \delta\bfu^{(0)} \cdot \left( \ddot{\bfu}^{(0)} + \xi \ddot{\bfu}^{(1)} \right) \dd\Omega \rev{\quad \forall \xi}. \label{eq:balanceeq}
\end{equation}
Obviously, the random variable $\xi$ occurs at multiple places. This, of course, is unsatisfying as it complicates any numerical solution.
Therefore, we investigate the equation further to arrive at deterministic expressions for the strain fields $\bfvarepsilon^0(t, \bfx)$ and $\bfvarepsilon^{(1)}(t, \bfx)$.
The main observation is that Equation~\eqref{eq:balanceeq} has to hold for any value of $\xi$.
Thus, one can recover the classical deterministic balance of linear momentum equation when Equation~\eqref{eq:balanceeq} is evaluated for $\xi = 0$. Consequently, the strain field $\bfvarepsilon^{(0)}(t)$ is determined by
\begin{equation}
\int_\Omega \delta\bfvarepsilon^{(0)} \cdot \bfsigma\vert_{\xi=0} \dd \Omega + \int_\Omega \delta\bfu^{(0)} \cdot \bfb^* \dd\Omega = \rho \int_\Omega \delta\bfu^{(0)} \cdot \ddot{\bfu}^{(0)} \dd\Omega.
\end{equation}
To derive an equation for the strain field $\bfvarepsilon^{(1)}$, we determine the tangent of Equation~\eqref{eq:balanceeq} with respect to the random variable at $\xi = 0$. Thus, the strain field $\bfvarepsilon^{(1)}$ can be derived by
\begin{equation}
\int_\Omega \delta\bfvarepsilon^{(0)} \cdot \frac{\dd \bfsigma}{\dd \xi}\Big\vert_{\xi = 0} \dd \Omega = \rho \int_\Omega \delta\bfu^{(0)} \cdot \ddot{\bfu}^{(1)} \dd\Omega
\end{equation}
with 
\begin{equation}
	\frac{\dd \bfsigma}{\dd \xi} \Big\vert_{\xi = 0} = \exp(-d^{(0)}) \left( (1- d^{(1)}) \dsE^{(0)} \cdot \bfvarepsilon^{(0)} + \dsE^{(0)} \cdot \bfvarepsilon^{(1)} \right).
\end{equation}

\subsubsection{Identification of terms: evolution of the internal variable}
\label{sec:InternalVariable}
As the internal variable $d(t)$ is expressed by the series expansion in Equation~\eqref{eq:dTSM}, one has to derive the corresponding evolution equations for the individual terms $d^{(0)}(t)$ and $d^{(1)}(t)$.
The evolution of the damage variable is given by the differential equation in Equation~\eqref{eq:damagediffeq}. This, of course, complicates the derivation of equations for $d^{(0)}(t)$ and $d^{(1)}(t)$. Nevertheless, a suitable procedure was presented in \cite{geisler_new_2024}.
The time-derivative of the Taylor series for the internal variable is given as
\begin{equation}
	\dot{d}^\mathrm{TSM}(t) = \dot{d}^{(0)}(t) + \xi \dot{d}^{(1)}(t)
\end{equation}
as the random variable is constant in time. 
Consequently, the Equation~\eqref{eq:damagediffeq} reads with all expansions as
\begin{equation}
    \dot{d}^{(0)} + \xi \dot{d}^{(1)} = \frac{1}{\eta} \exp(-(d^{(0)}+\xi d^{(1)})) \frac{1}{2} (\bfvarepsilon^{(0)} + \xi \bfvarepsilon^{(1)}) \cdot (\dsE^{(0)} (1+\xi)) \cdot (\bfvarepsilon^{(0)} + \xi \bfvarepsilon^{(1)}). \label{eq:expandeddamagediffeq}
\end{equation}
The evolution equations for $d^{(0)}(t)$ and $d^{(1)}(t)$ are found similarly to Section~\ref{sec:BalanceEquation} by derivative matching.
The evaluation of Equation~\eqref{eq:expandeddamagediffeq} at $\xi = 0$ yields the standard deterministic evolution equation \rev{which is analogous to}
\begin{equation}
    \dot{d}^{(0)}(t) = \frac{1}{\eta} \exp(-d^{(0)}) \left( \frac{1}{2} \bfvarepsilon^{(0)} \cdot \dsE{(0)} \cdot \bfvarepsilon^{(0)} \right).
\end{equation}
The evolution equation for the field $d^{(1)}(t)$ is most easily found by differentiation of both sides of Equation~\eqref{eq:expandeddamagediffeq} with respect to $\xi$ and evaluation at $\xi = 0$ as
\begin{align}
    \dot{d}^{(1)}(t) = \frac{1}{\eta} \exp(-d^{(0)}) \left( (1-d^{(1)}(t)) \, \bfvarepsilon^{(0)} \cdot \dsE^{(0)} \cdot \bfvarepsilon^{(0)} + 2 \bfvarepsilon^{(1)} \cdot \dsE^{(0)} \cdot \bfvarepsilon^{(0)} \right).
\end{align}
Thus, we arrive at evolution equations for the terms $d^{(0)}(t)$ and $d^{(1)}(t)$. This is only reasonable as the internal variable $d(t)$ is given as a differential equation.

\subsection{Uncertainty quantification}
\label{sec:UncertaintyQuantification}
The series for the displacement and damage field as given in Section~\ref{sec:SeparatedRepresentation} can be evaluated for any value of the random variable $\xi$. As a linearization of the fields around $\xi = 0$, the approximation is best around the expectation of the random variable, i.e., $\xi = 0$.
For engineering applications, one is typically interested in the stochastic characteristics, e.g., expectation and standard deviation of various quantities of interest. In this section, we present the calculation of the expectation and standard deviation for the damage $d$, the damage function $f$, the stress $\bfsigma$ and reaction force $\bfF$. Only the second moment of the random variable $\langle \xi^2 \rangle$ \rev{needs} to be known. Thus, the approach can be used for a large class of probability density functions of the random variable.

\subsubsection{Internal variable}
The damage field is approximated by the series expansion in Equation~\eqref{eq:dTSM}. Naturally, the expectation results as
\begin{equation}
    \langle d \rangle = d^{(0)}
\end{equation}
and the standard deviation is given with Equation as
\begin{equation}
    \textrm{Std}(d) = \sqrt{\langle \xi^2 \rangle} d^{(1)}(t)
\end{equation}

\subsubsection{Damage function} 
Similar to before, we linearize the algebraic equation for the damage function $f$ as
\begin{equation}
	f^\mathrm{TSM}(t, \xi) = f^{(0)} + \xi f^{(1)} = f(t, \xi = 0) + \xi \frac{\dd f}{\dd \xi} \Big\vert_{\xi = 0} \approx f(t, \xi).
\end{equation}
The linearization allows us to express the expectation as
\begin{equation}
	\langle f \rangle = f^{(0)} = \exp(-d^{(0)})
\end{equation}
and the standard deviation as
\begin{equation}
	\textrm{Std}(f) = \sqrt{\langle \xi^2 \rangle} f^{(1)}(t)
\end{equation}

\subsubsection{Stress}
The stress as given by Equation~\eqref{eq:stresseq} is nonlinearly dependent on the random variable.
To quantify expectation and standard deviation, we linearize the stress for each component $a$ as
\begin{equation}
	\sigma_a = \sigma^{(0)}_a + \xi \sigma^{(1)}_a
\end{equation}
with
\begin{equation}
	\sigma^{(0)}_a = \sigma_a \vert_{\xi = 0} = \exp(-d^{(0)}) \dsE^{(0)}_{ab} \cdot \varepsilon_b^{(0)}
\end{equation}
and 
\begin{equation}
	\sigma_a^{(1)} = \frac{\dd \sigma_a}{\dd \xi} \Big\vert_{\xi = 0} = \exp(-d^{(0)}) ((1- d^{(1)}) \dsE^{(0)}_{ab} \cdot \varepsilon^{(0)}_b + \dsE^{(0)}_{ac} \varepsilon^{(1)}_c).
\end{equation}
Therefore, the expectation results as
\begin{equation}
    \langle \bfsigma \rangle = \bfsigma^{(0)}
\end{equation}
and the standard deviation as
\begin{equation}
    \textrm{Std}(\sigma_a) = \sqrt{\langle \xi^2 \rangle} \bfsigma_a^{(1)}.
\end{equation}

\subsubsection{Reaction force}
\rev{The} reaction force as a global quantity gives important hints about the evolution of damage. It is given in a finite element context as
\begin{equation}
    \bfF = \int_\Omega \bfB^T \cdot \bfsigma + \rho \rev{\bfN \cdot} \ddot{\bfu} \,\mathrm{d}V,
\end{equation} which is evaluated at the non-zero Dirichlet boundary.
The expectation is given as
\begin{equation}
	\langle F_a \rangle = \int_\Omega \left( \bfB^T \cdot \bfsigma^{(0)} + \rho \rev{\bfN \cdot} \ddot{\bfu}^{(0)} \right)_a \,\mathrm{d}V
\end{equation}
for each component $a$.
The standard deviation results as
\begin{equation}
	\textrm{Std}(F_a) = \sqrt{\langle \xi^2 \rangle} \int_\Omega \left( \bfB^T \cdot \bfsigma^{(1)} + \rho \rev{\bfN \cdot} \ddot{\bfu}^{(1)} \right)_a \,\mathrm{d}V
\end{equation}
for each component $a$.

\section{\rev{Weakly-intrusive implementation}}
\label{sec:ImplAbaqSec}
In this section, we present the implementation of the time-separated stochastic mechanics (TSM) in the finite element software Abaqus.
This allows the adoption of the TSM for a wider user base as Abaqus allows for a pleasant handling of complicated geometries, boundary conditions and meshes.

The TSM does not need any modification as it is \rev{weakly}-intrusive in nature. In fact, we do not require any coupling to external software. 
Here, we chose the solver Abaqus/Explicit as the characteristic time $\tau_0$ is typically in the order of milliseconds and therefore a high time-resolution is necessary. Nevertheless, a similar implementation in Abaqus/Standard or other finite element software is straight-forward.

The governing equations to be solved are presented in Section~\ref{sec:GovEq}. The specific usage of the Abaqus subroutines is discussed in Section~\ref{sec:ImplementationAbaqus}.

\subsection{Governing equations}
\label{sec:GovEq}
Due to the linear series expansion for the strain and the internal variable, two terms each have to be determined.
These are the terms $\bfvarepsilon^{(0)}(t)$ and $\bfvarepsilon^{(1)}(t)$ for the strain and $d^{(0)}(t)$ and $d^{(1)}(t)$ for the damage variable, c.f., Section~\ref{sec:SeparatedRepresentation}.

The governing equations to calculate these terms are given in Sections~\ref{sec:BalanceEquation} and \ref{sec:InternalVariable} and are recalled here for convenience.
The balance of linear momentum and evolution equation which allow to calculate $\bfvarepsilon^{(0)}(t)$ and $d^{(0)}(t)$ are given by
\begin{align}
\int_\Omega \delta\bfvarepsilon^{(0)} \cdot \exp(-d^{(0)}) \dsE^{(0)} \cdot \bfvarepsilon^{(0)} \dd \Omega &+ \int_\Omega \delta\bfu^{(0)} \cdot \bfb^* \dd\Omega = \rho \int_\Omega \delta\bfu^{(0)} \cdot \ddot{\bfu}^{(0)} \dd\Omega \nonumber \\
 \dot{d}^{(0)}(t) &= \frac{1}{\eta} \exp(-d^{(0)}) \left( \frac{1}{2} \bfvarepsilon^{(0)} \cdot \dsE^{(0)} \cdot \bfvarepsilon^{(0)} \right). \label{eq:System0}
\end{align} 
We refer to the system of equations above as the zeroth-order equation set as it is used to determine the terms of zeroth order of the respective Taylor series.
The terms $\bfvarepsilon^{(1)}(t)$ and $d^{(1)}(t)$ are determined by the set of equations
\begin{align}
	\int_\Omega \delta\bfvarepsilon^{(0)} \cdot \exp(-d^{(0)}) &\left( (1- d^{(1)}) \dsE^0 \cdot \bfvarepsilon^{(0)} + \dsE^{(0)} \cdot \bfvarepsilon^{(1)} \right) \dd \Omega = \rho \int_\Omega \delta\bfu^{(0)} \cdot \ddot{\bfu}^{(1)} \dd\Omega \nonumber \\
	\dot{d}^{(1)}(t) &= \frac{1}{\eta} \exp(-d^{(0)}) \left( (1-d^{(1)}(t)) \, \bfvarepsilon^{(0)} \cdot \dsE^{(0)} \cdot \bfvarepsilon^{(0)} + 2 \bfvarepsilon^{(1)} \cdot \dsE^{(0)} \cdot \bfvarepsilon^{(0)} \right) \label{eq:System1}
\end{align}
which we similarly refer to as first-order equation set.
The \rev{set of} equation as given by Equations~\eqref{eq:System0} and \eqref{eq:System1} can be solved in a staggered manner. \rev{This means, instead of a monolithic approach, which increases the size of the equation system, the equations are handled individually and sequentially.}
It may be remarked, that the second equation set needs the solution of $d^{(0)}(t)$ and $\bfvarepsilon^{(0)}(t)$ as prerequisite to be solved.  

\subsection{Implementation scheme \rev{in detail}}
\label{sec:ImplementationAbaqus}
The equation sets~\eqref{eq:System0} and \eqref{eq:System1} can be easily implemented by a custom user-defined material model (\texttt{vumat}) in Abaqus/Explicit. The internal variables $d^{(0)}(t)$ and $d^{(1)}(t)$ can be stored via a state variable.
In Abaqus/Explicit, the user-defined material model shall return the stress for the particular timestep.
The stress at timestep $n+1$ for the zeroth-order equation set is given by
\begin{equation}
	\bfsigma^{(0)}_{n+1} = \exp(-d^{(0)}_n) \, \dsE^0 \cdot \bfvarepsilon^0_n
\end{equation} 
and for the first-order equation set as
\begin{equation}
	\bfsigma^{(1)}_{n+1} = \exp(-d^{(0)}_n) \left( (1- d^{(1)}_n) \, \dsE^{(0)} \cdot \bfvarepsilon^{(0)}_n + \dsE^{(0)} \cdot \bfvarepsilon^{(1)}_n \right).
\end{equation} 
As the first-order simulation needs access to the results for $\bfvarepsilon^{(0)}, d^{(0)}$ of the zeroth-order simulation, a suitable exchange interface has to be implemented. Abaqus allows a coupling of different simulations via Co-Simulation. However, this coupling scheme was designed with multi-physics applications as thermo-mechanical coupling or fluid-structure interaction in mind. While being a valuable tool, here, we resort to a simpler implementation scheme.
In this work, the results of the zeroth-order simulation are exported in a user-readable file at each timestep. These files are read in by the first-order simulation and handled as an external field. The user-defined subroutines \texttt{vexternaldb} and \texttt{vusdfld} allow for an efficient implementation of this scheme.
Of course, file operations are an inherently slow process as the communication with the operating system has to be performed. This can be a limiting factor in explicit simulations as the number of timesteps is rather high resulting in many file operations. 
As a mitigation strategy, we perform the file writing/reading process only for a subset of all timesteps and assume a constant evolution in between. Of course, this procedure is inherently linked to multirate formulations \cite{gear_multirate_1984}.
As a heuristic rule, we choose to perform the exchange process every 1000 time steps. \rev{This number depends on the dynamics of the problem at hand and thus, must be adjusted.}
The uncertainty quantification following Section~\ref{sec:UncertaintyQuantification} is carried out in a post-processing step after both simulations finished. For the post-processing the Python Scripting Interface is used.
Result of the calculation are the expectation and standard deviation of all fields, e.g., stress and internal variable, over time. These are written in the output database (\texttt{.odb}) of the job. Additionally, the expectation and standard deviation of the reaction force are computed. 

\begin{figure}
\centering
\includegraphics[scale=0.5]{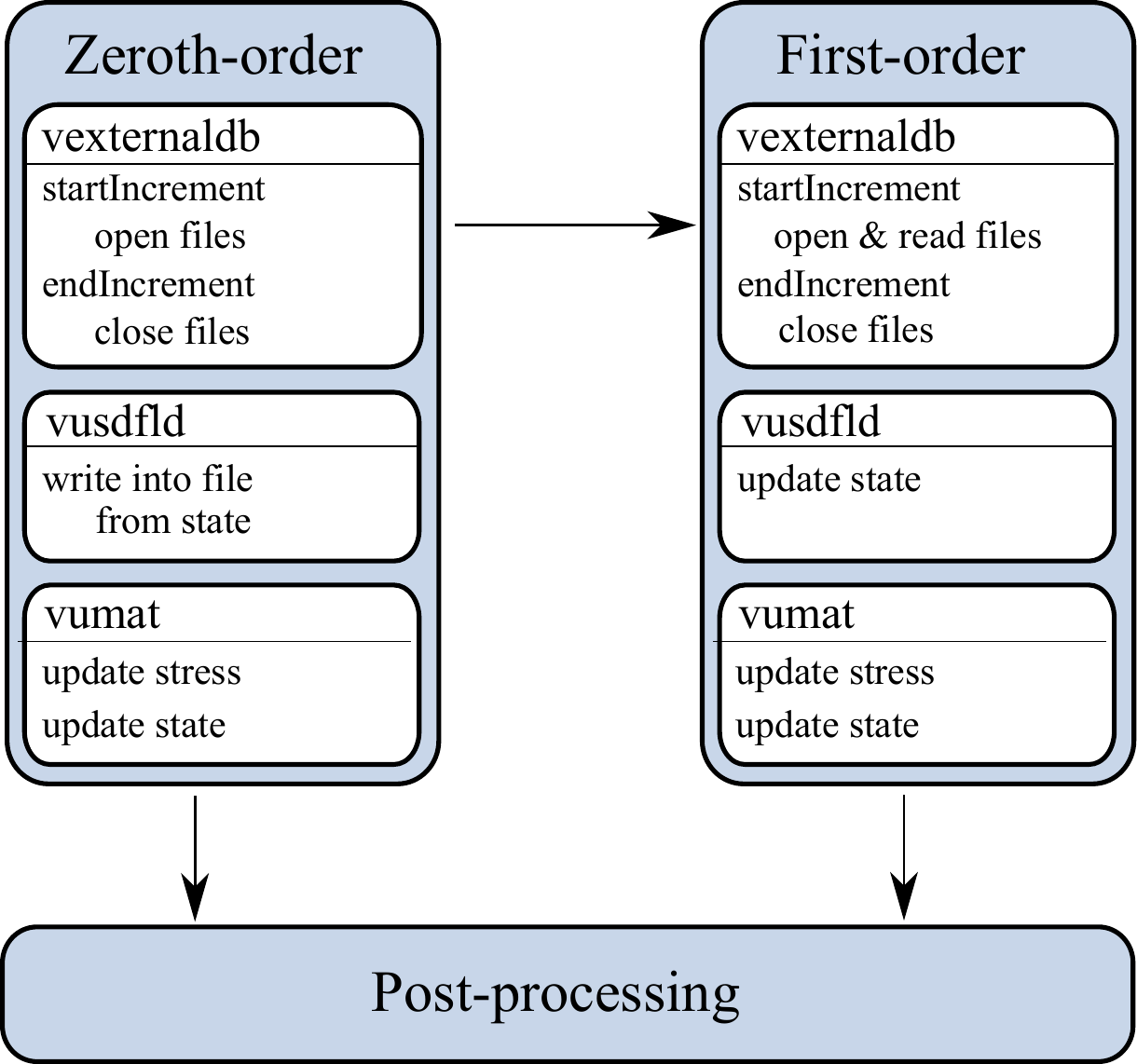}
\caption{Implementation scheme of the time-separated stochastic mechanics in Abaqus. The user-defined subroutines \texttt{vexternaldb}, \texttt{vusdfld}, \texttt{vumat} enable the coupling of the simulations.}
\label{fig:ImplementationScheme}
\end{figure}

The overall implementation scheme is presented in Figure~\ref{fig:ImplementationScheme}. In this figure the individual operations as file opening, writing and closing are sorted in the corresponding user subroutines.
 
\section{Numerical results}
\label{sec:NumericalResults}
As the computational cost of the Monte Carlo method is often prohibitively large for three-dimensional problems, we focus on three-dimensional structures \rev{with vanishing thickness}. In particular, we investigate a double notched specimen and a plate with a hole. For these problems, we compare the results between our approach and reference Monte Carlo simulations. For the Monte Carlo simulation, 500 sampling points are chosen. For each sampling point, the system is simulated for the whole time-domain albeit with different realizations of the material parameters. In contrast, the time-separated stochastic mechanics only needs two deterministic simulations. 

For both boundary value problems, the same material parameters are chosen. The mean value of the Lam\'e parameters are $\lambda = \SI{1}{GPa}$ and $\mu = \SI{0.8}{GPa}$. The density is set as $\rho = \SI{1000}{kg/m^3}$. The standard deviation of the \rev{random variable of the elasticity tensor}, c.f. Equation~\eqref{eq:Erand}, is chosen as $10\%$. 

\subsection{Double-notched specimen}
\begin{figure}
	\centering
	\includegraphics{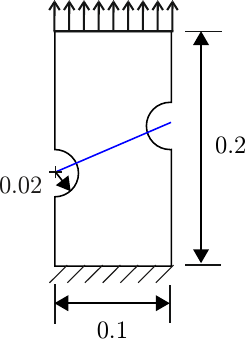}
	\caption{Geometry of the double-notched specimen with the boundary conditions. Additionally, the orientation of a line for closer investigation of the results is indicated. All lengths are given in the unit $m$.}
	\label{fig:DNSGeometry}
\end{figure}

The first investigated boundary value problem is a double notched specimen. The geometry and boundary conditions are depicted in Figure~\ref{fig:DNSGeometry}. The specimen is discretized with 646 brick-elements and the time-discretization is set as $\Delta t = \SI{1.6}{\micro s}$. A time-proportional displacement is applied on the top boundary with the maximal displacement $\SI{0.01}{m}$ at $t = \SI{1}{s}$. The viscosity is set to $\eta = \SI{1}{GPa.s}$.
The evolution of the damage is visible in Table~\ref{tab:DNSfovertime}. In this table, the expectation and standard deviation of the damage function $f$ are presented for three different time steps. The damage starts at the boundary of the two notches and spreads through the structure.
After $t = \SI{0.5}{s}$, a damaged zone between both notches is clearly visible. At $t = \SI{0.66}{s}$, nearly the whole plate is damaged. 
As the results for the expectation $\langle f \rangle$ of TSM and MC are indistinguishable, only the result of the TSM is presented here. The highest standard deviation of the damage variable $f$ is reported for all time steps at the highly damaged zones around the notches. The standard deviation of the damage function $f$ grows with the evolution of the damage. This behavior is correctly reflected by the TSM results. However, we report differences for the standard deviation at the highly damaged elements. This typically occurs for elements with $\langle f \rangle \leq 0.5$. Nevertheless, for the rest of the specimen the evolution of the standard deviation shows only very small deviations from the MC results.

\begin{figure}
	\centering
	\begin{subfigure}[b]{0.32\textwidth}
		\centering
		\includegraphics[width=\textwidth]{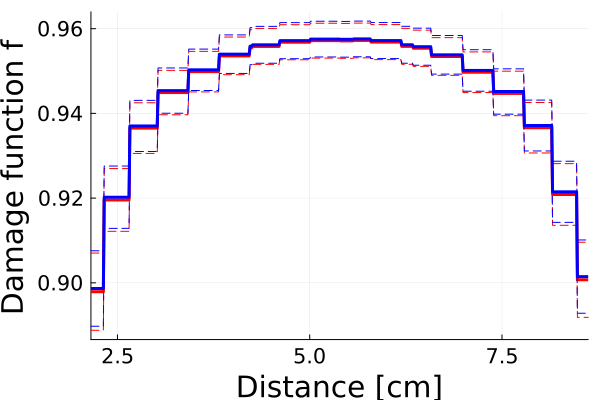}
		\caption{0.33 s}
	\end{subfigure}
	\hfill
	\begin{subfigure}[b]{0.32\textwidth}
		\centering
		\includegraphics[width=\textwidth]{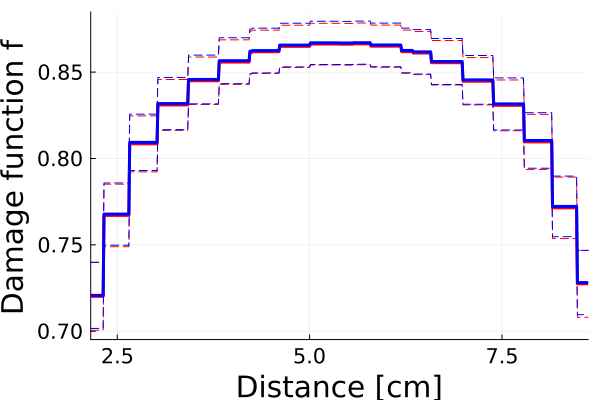}
		\caption{0.5 s}
	\end{subfigure}
	\hfill
	\begin{subfigure}[b]{0.32\textwidth}
		\centering
		\includegraphics[width=\textwidth]{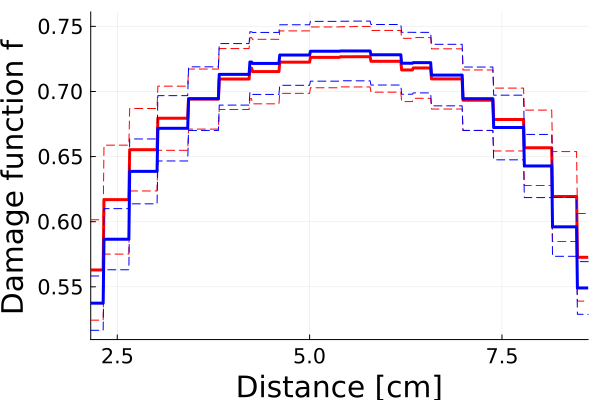}
		\caption{0.66 s}
	\end{subfigure}
	\caption{Results for the damage function $f$ along the line as indicated in Figure~\ref{fig:DNSGeometry} at three time points. The expectation $\langle f \rangle$ is given by a solid line, the results for $\langle f \rangle \pm \textrm{Std}(f)$ by dashed lines. The results of the TSM are given in blue and the results of MC in red.}
	\label{fig:DNSf}
\end{figure}

To further investigate the quality of approximation, the results of TSM and MC for the damage function $f$ are compared along a line between the two notches. This line is indicated in blue in Figure~\ref{fig:DNSGeometry}. The results for three time points are presented in Figure~\ref{fig:DNSf}. Here, the expectation is presented as a solid line, the values for $\langle f \rangle \pm \textrm{Std}(f)$ are indicated by dashed lines. The results for the TSM are given in blue and the results of the Monte Carlo method in red. Only the average of the values in each element is reported, i.e., no interpolation is done, such that a step-shaped pattern results. For $t = \SI{0.33}{s}$ and $t = \SI{0.5}{s}$ the results are visually identical. The loss in stiffness is clearly visible at the borders of the figure. For $t = \SI{0.66}{s}$, differences at the highly damaged zones near the two notches are visible as discussed before. 

The expectation and standard deviation of the norm of the stress at $t = \SI{0.5}{s}$ is presented in Table~\ref{tb:DNSStress05}. The differences between TSM and MC are barely visible. In fact, the expectation is identical whereas a difference in the standard deviation of the damage is visible at the highly damaged zone on the boundary of the two notches. This fits with our observation that the TSM provides a surprisingly good approximation of expectation and standard deviation especially during the onset of the damage evolution.

\begin{table}
	\caption{Evolution of the damage function $f$ for three time points. The results for the expectation $\langle f \rangle$ are visually identical for TSM and MC. The results for the standard deviation $\textrm{Std}(f)$ are distinguished between TSM and MC.}
	\begin{tabular}{|p{2cm}|p{3cm}p{3cm}p{3cm}p{1cm}}
		\hline $t$ & 0.33s & 0.5s & 0.66s & \\
		\hline \centered{$\langle f \rangle$} & \centered{\includegraphics[scale=0.25, trim=22cm 3cm 22cm 3cm,clip]{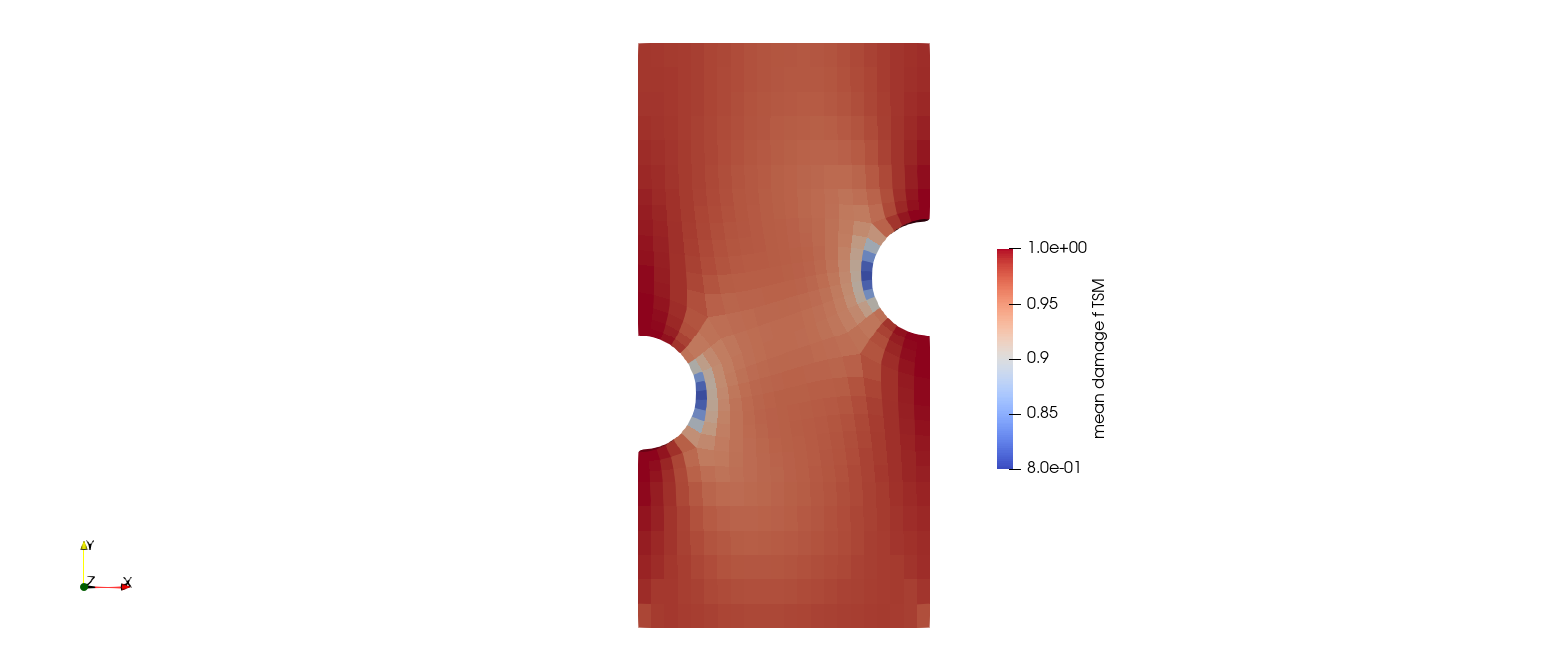}} & \centered{\includegraphics[scale=0.25, trim=22cm 3cm 22cm 3cm,clip]{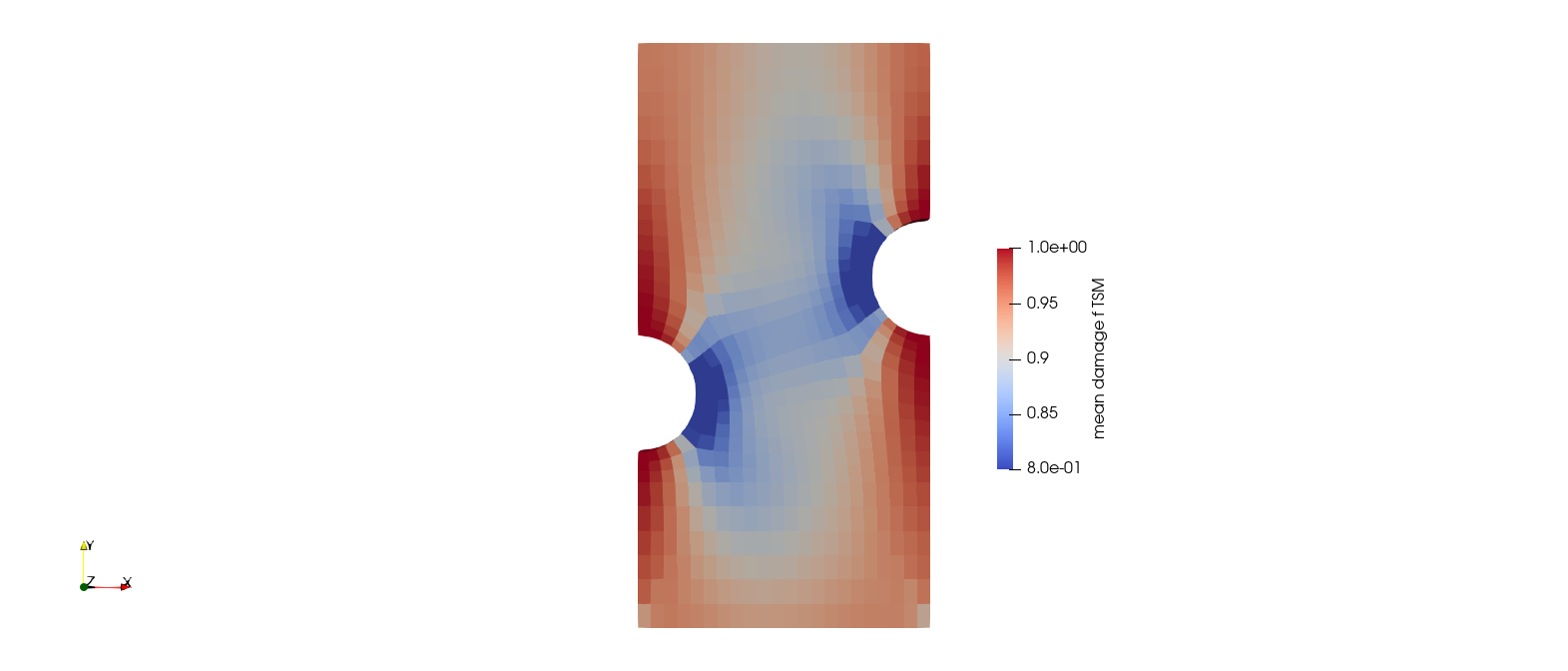}} & \centered{\includegraphics[scale=0.25, trim=22cm 3cm 22cm 3cm,clip]{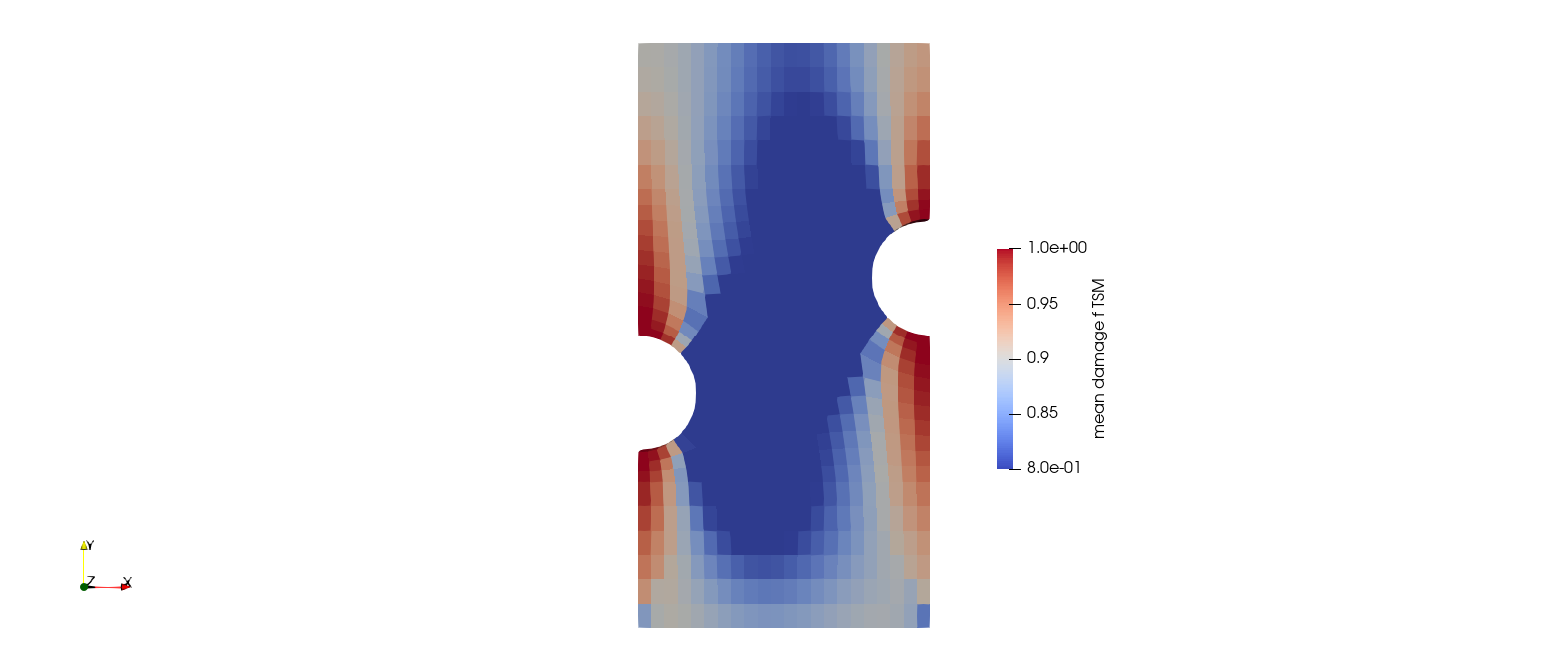}} & \centered{\includegraphics[scale=0.4, trim=34cm 7cm 17cm 7cm,clip]{plots/Plate/mean_f_TSM_66.png}}\\ 
		
		\hline \centered{$\textrm{Std}(f)$ \\ by MC} & \centered{\includegraphics[scale=0.25, trim=22cm 3cm 22cm 3cm,clip]{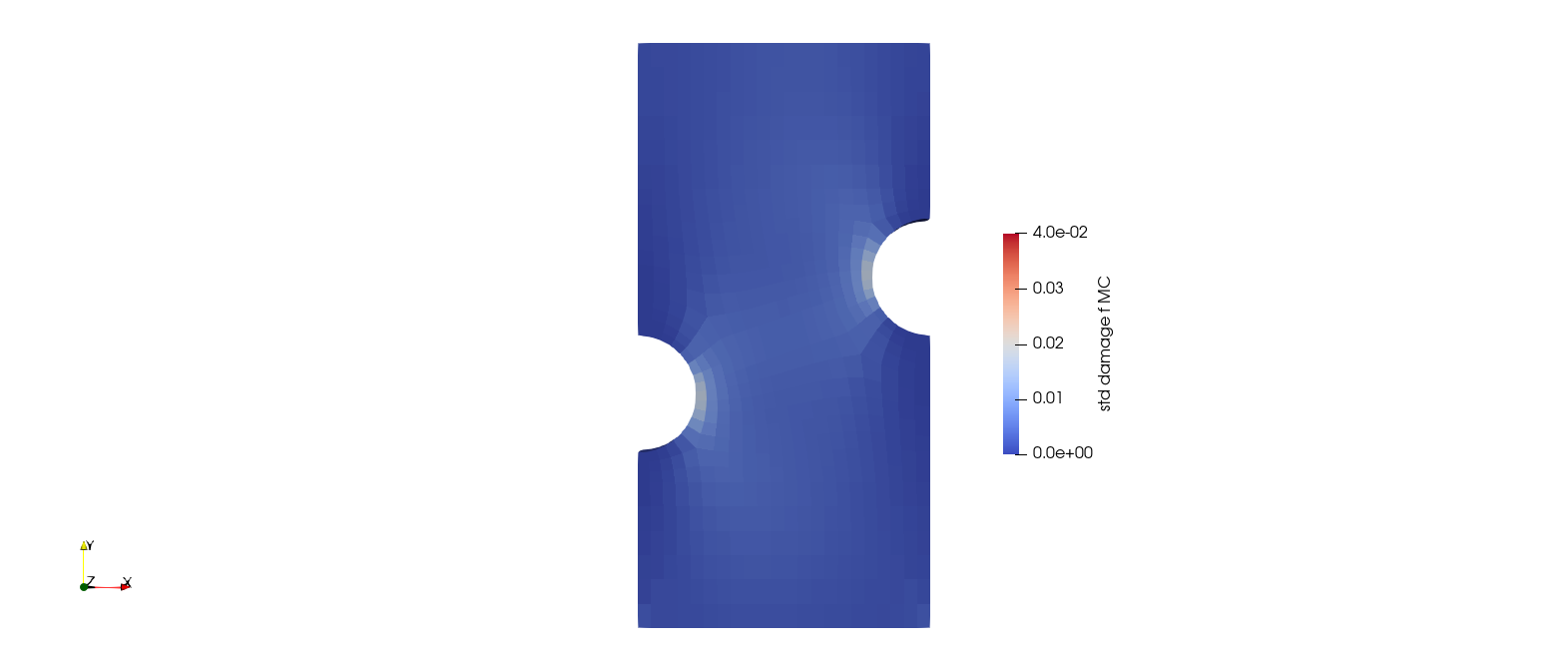}} & \centered{\includegraphics[scale=0.25, trim=22cm 3cm 22cm 3cm,clip]{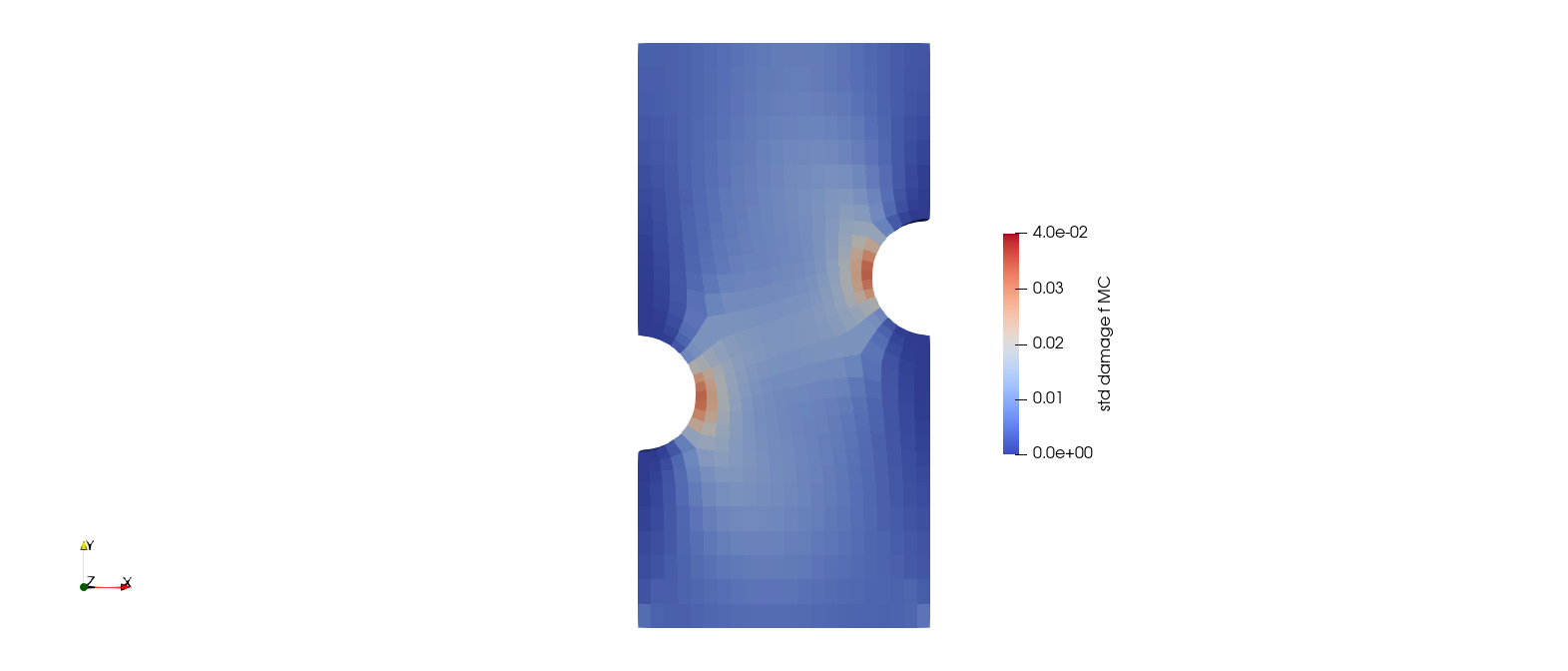}} & \centered{\includegraphics[scale=0.25, trim=22cm 3cm 22cm 3cm,clip]{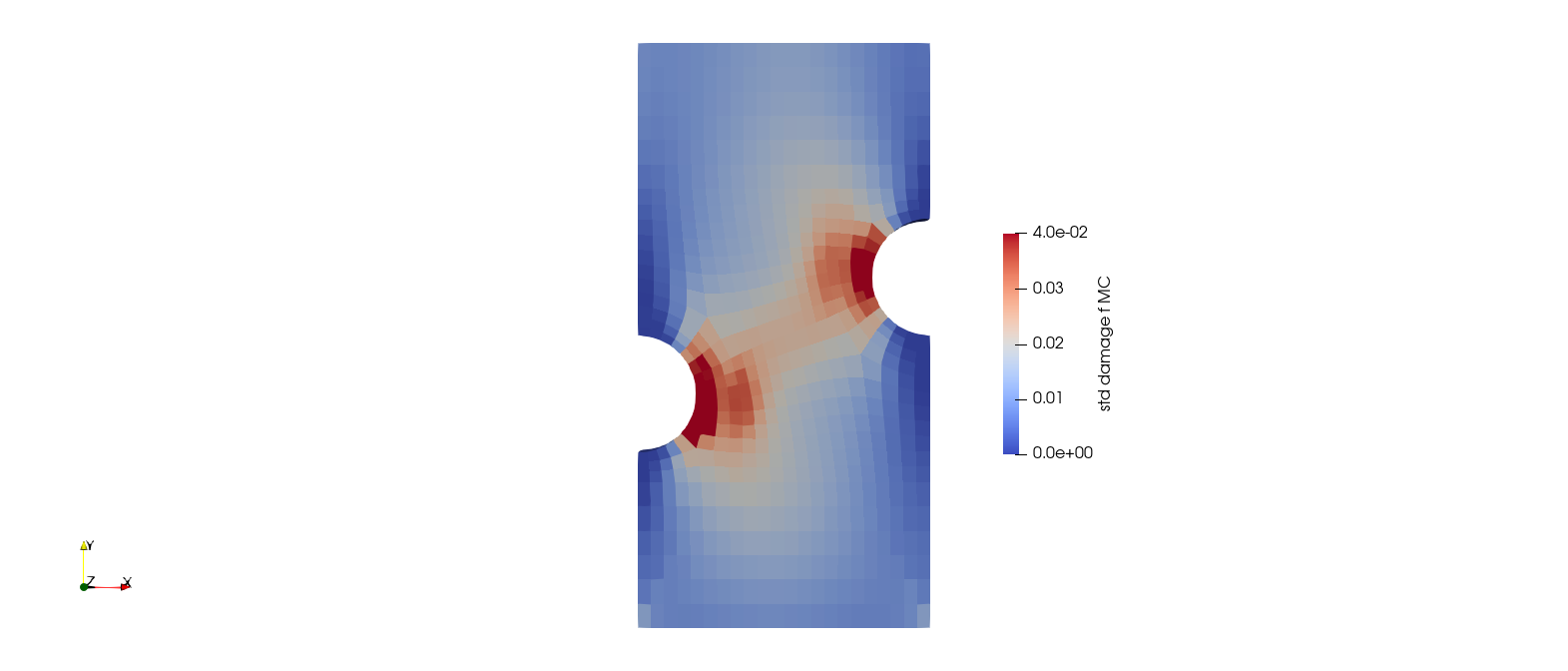}} & \centered{\includegraphics[scale=0.4, trim=34cm 7cm 16.7cm 7cm,clip]{plots/Plate/std_f_MC_66.png}}\\
		
		\hline \centered{$\textrm{Std}(f)$ \\ by TSM} & \centered{\includegraphics[scale=0.25, trim=22cm 3cm 22cm 3cm,clip]{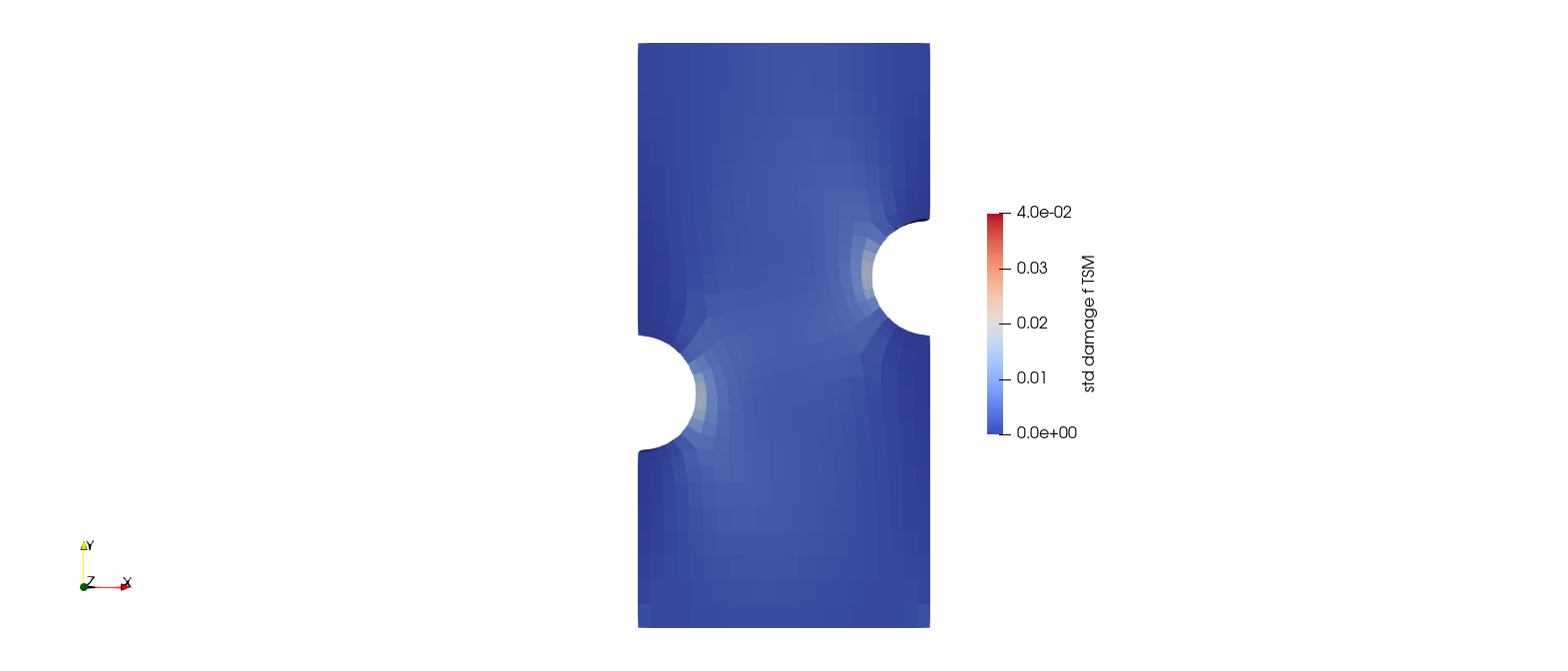}} & \centered{\includegraphics[scale=0.25, trim=22cm 3cm 22cm 3cm,clip]{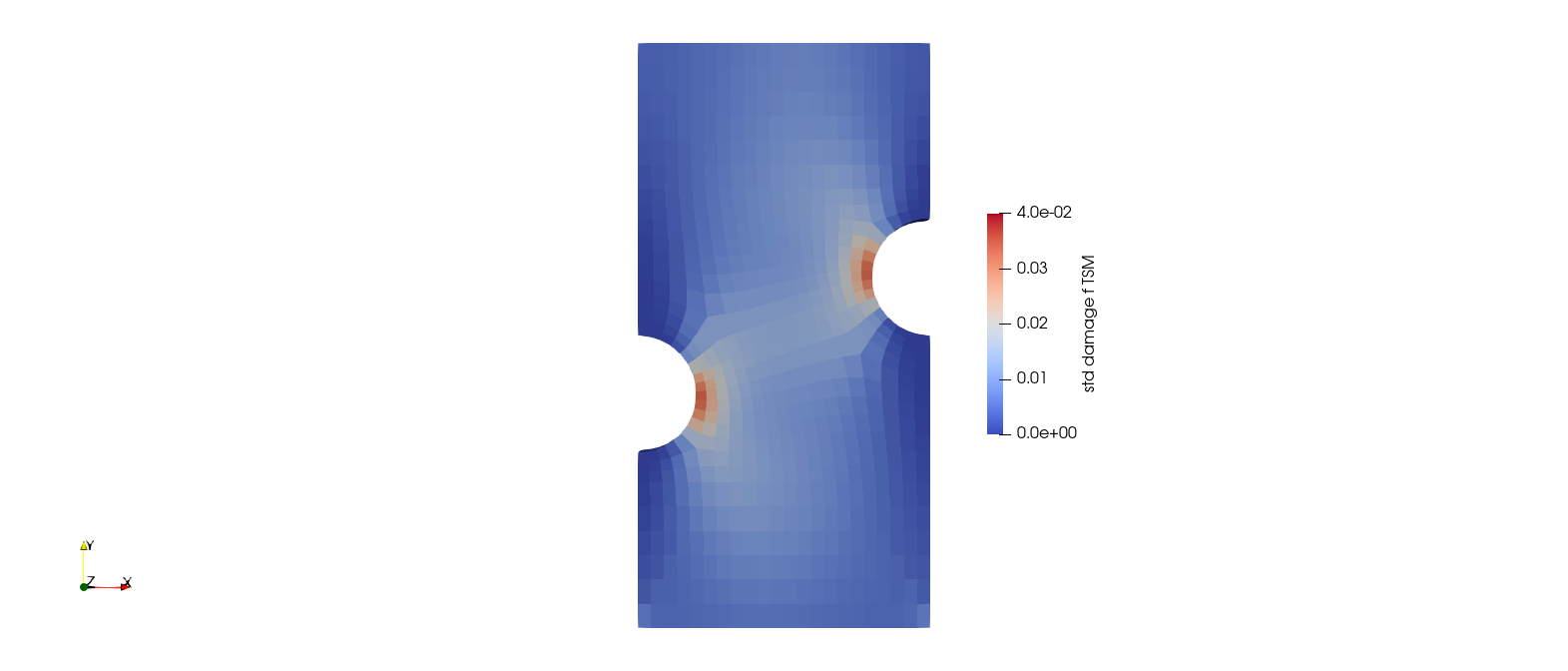}} & \centered{\includegraphics[scale=0.25, trim=22cm 3cm 22cm 3cm,clip]{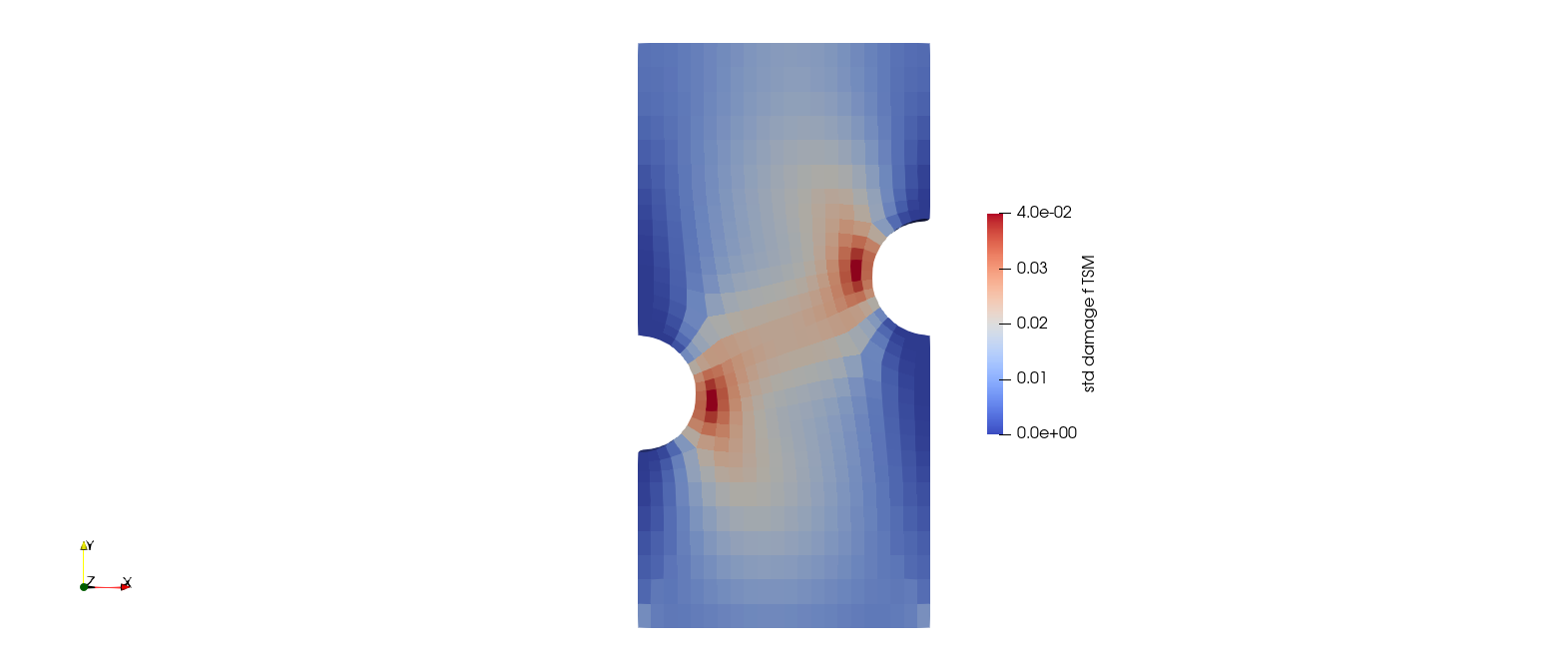}} & \centered{\includegraphics[scale=0.4, trim=33.5cm 7cm 17.3cm 7cm,clip]{plots/Plate/std_f_TSM_66.png}}\\
		\hline 
	\end{tabular} 
	\label{tab:DNSfovertime}
\end{table}

\begin{table}
	\caption{Results for the stress in the specimen at $t = \SI{0.5}{s}$.}
	\begin{tabular}{p{2cm}|p{4cm}|p{4cm}|p{2cm}}
		\hline 
		& TSM & MC & \\ \hline
		\centered{$||\langle \bfsigma \rangle||$} & \centered{\includegraphics[scale=0.2, trim=30cm 3cm 30cm 3cm, clip]{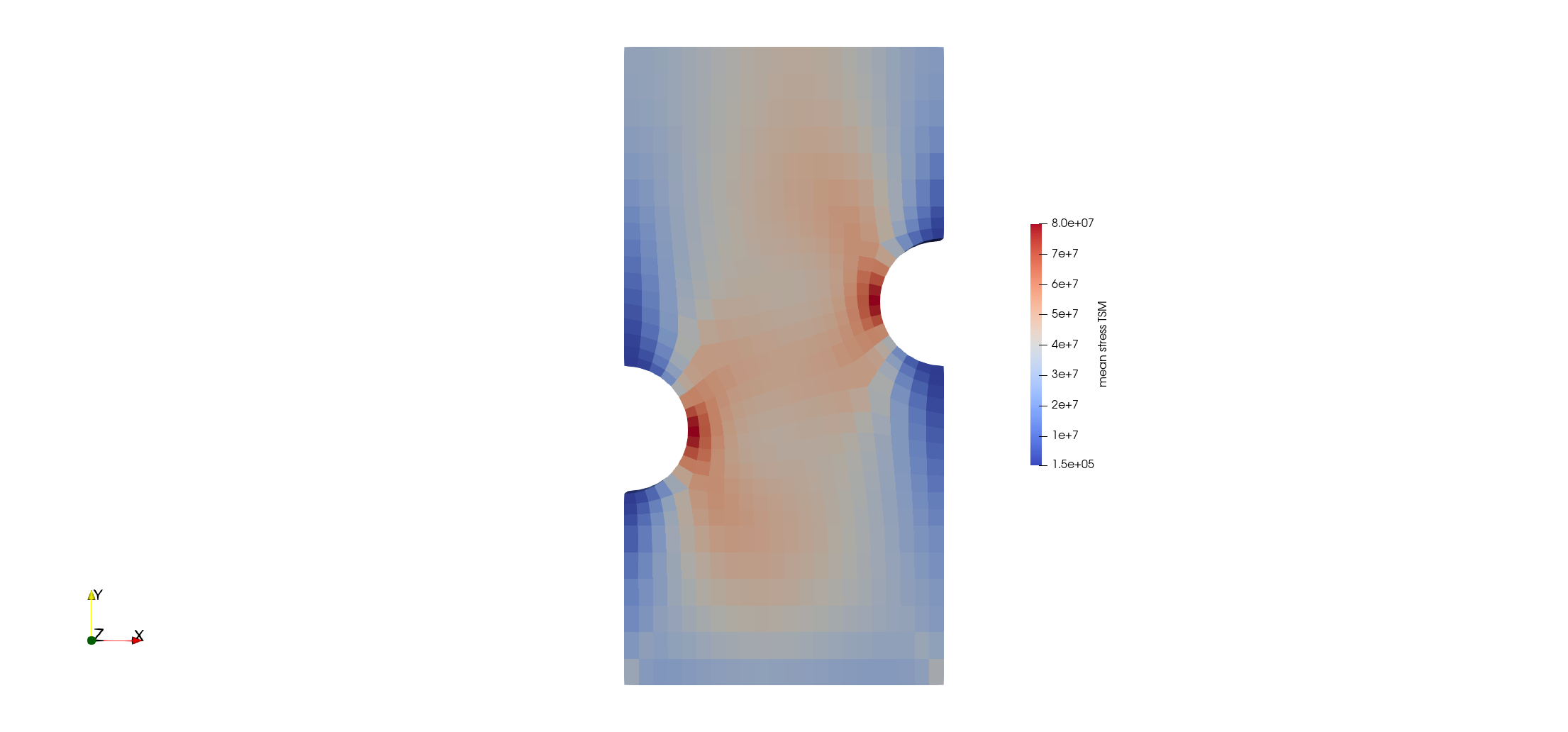}} & \centered{\includegraphics[scale=0.2, trim=30cm 3cm 30cm 3cm, clip]{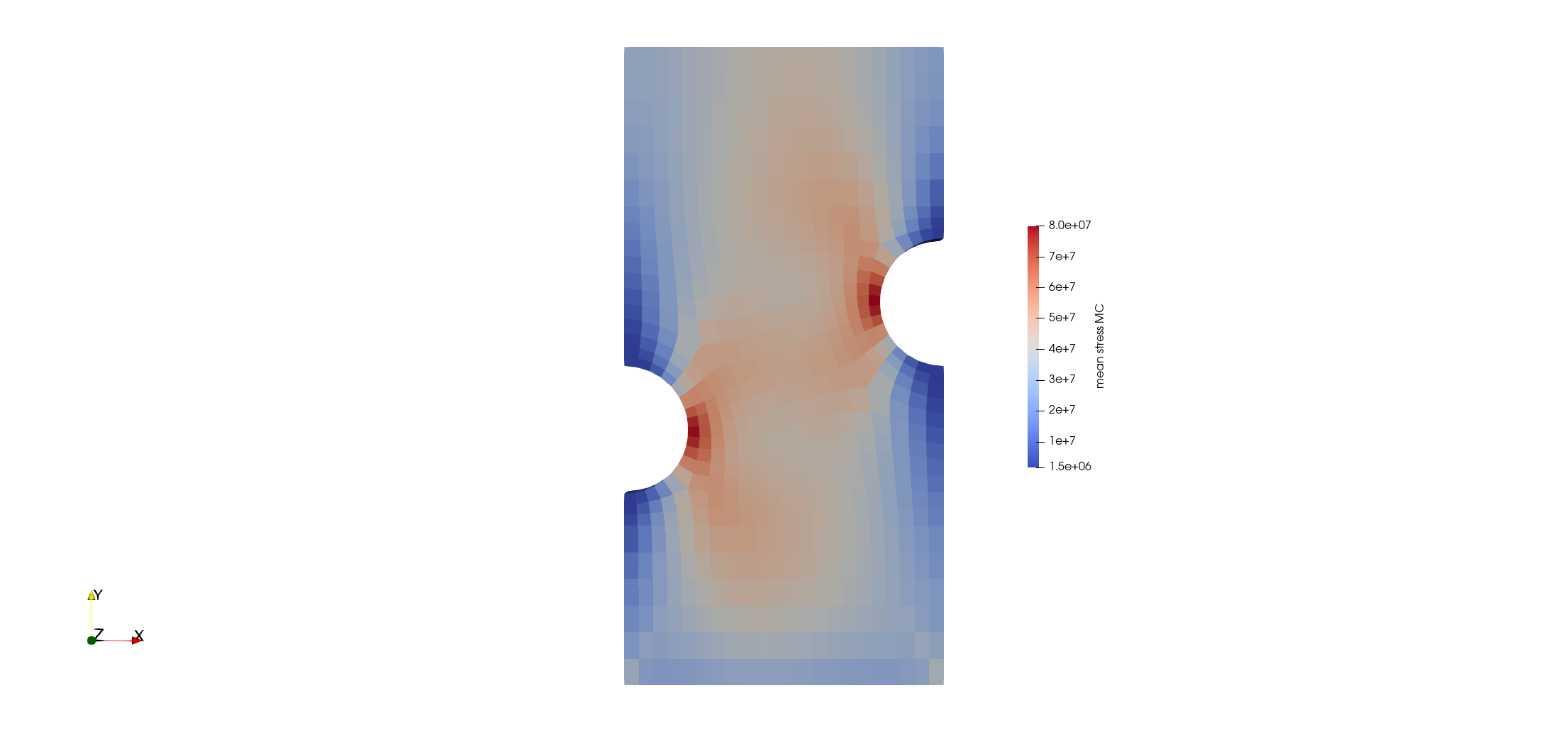}} & \centered{\includegraphics[scale=0.3, trim=50cm 8cm 23.5cm 8cm, clip]{plots/Plate/mean_s_TSM_50.png}}\\
		\centered{$||\textrm{Std}(\bfsigma)||$} & \centered{\includegraphics[scale=0.2, trim=30cm 3cm 30cm 3cm, clip]{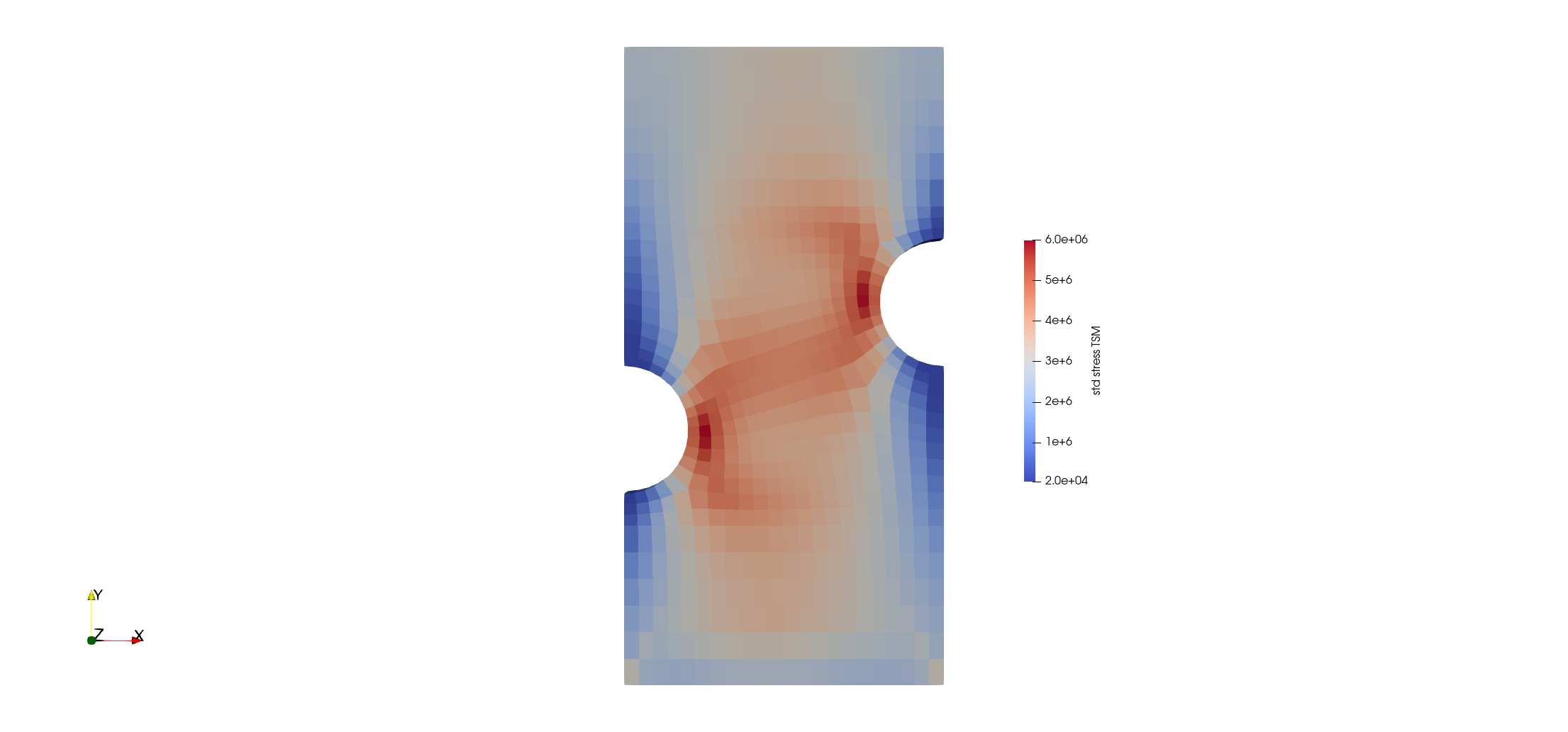}} & \centered{\includegraphics[scale=0.2, trim=30cm 3cm 30cm 3cm, clip]{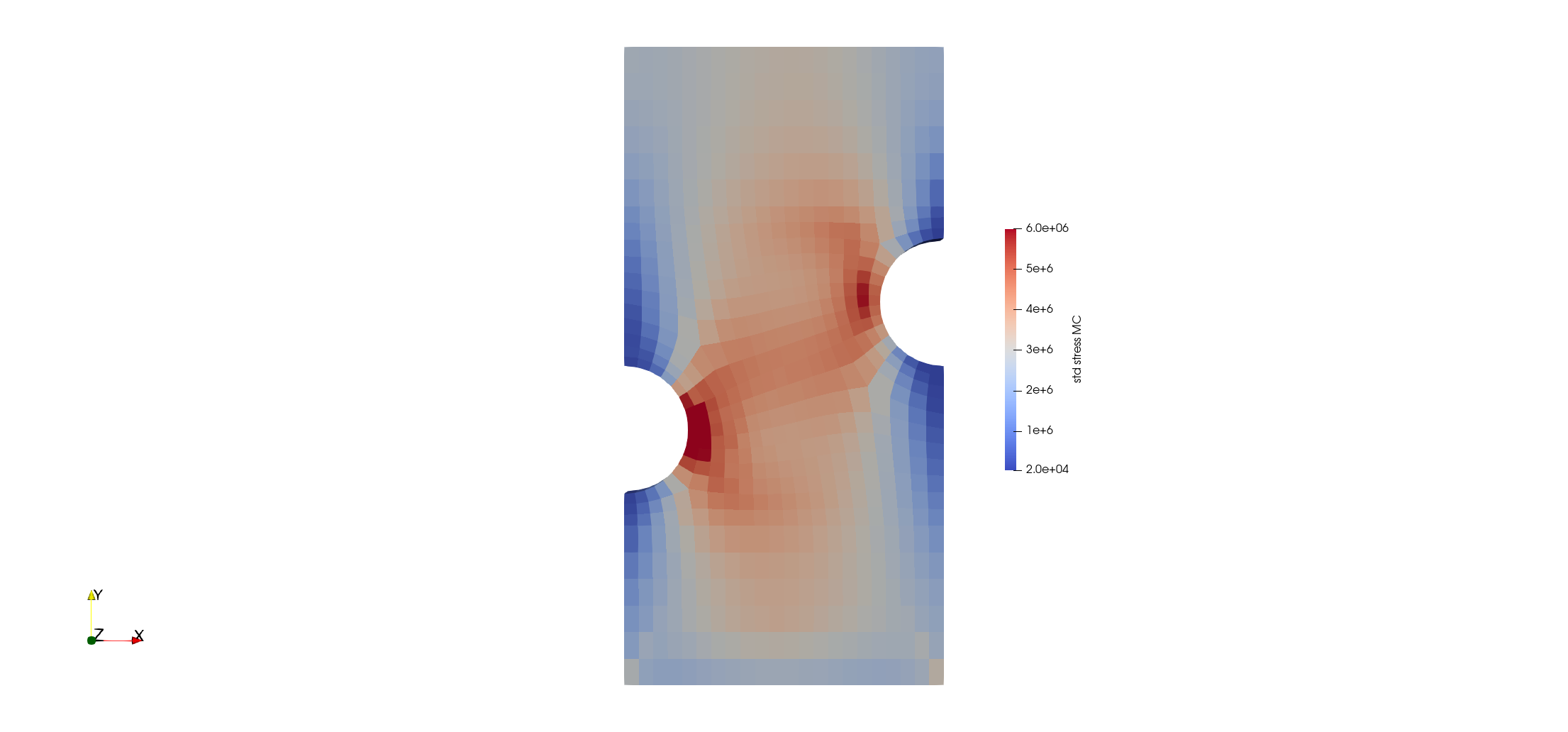}} & \centered{\includegraphics[scale=0.3, trim=50.2cm 8cm 23.7cm 8cm, clip]{plots/Plate/std_s_TSM_50.png}}\\
	\end{tabular}
	\label{tb:DNSStress05}
\end{table}

\subsection{Plate with hole}
\begin{figure}
	\centering
	\includegraphics{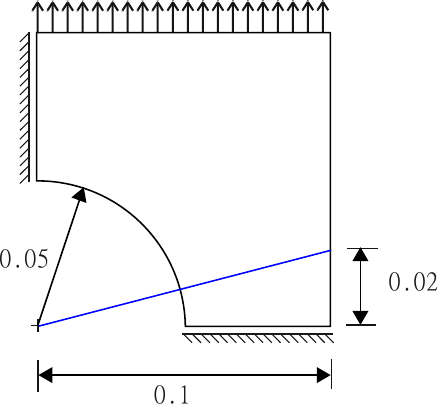}
	\caption{Geometry of the plate with hole with the boundary conditions. Additionally, the orientation of a line for closer investigation of the results is indicated. All lengths are given in the unit $m$.}
	\label{fig:PlateGeometry}
\end{figure}

The second boundary value problem is a plate with a hole. Due to symmetry, we only discretize one quarter and apply corresponding symmetry boundary conditions. The plate is discretized by 594 brick-elements. The time-discretization is chosen as $\Delta t = \SI{1.8}{\micro s}$. The geometry and the boundary conditions are presented in Figure~\ref{fig:PlateGeometry}. A time-proportional displacement is applied on the top boundary with a maximum displacement of $\SI{1}{mm}$ at $t = \SI{1}{s}$. The viscosity is set to $\eta = \SI{5}{GPa.s}$ for this problem.

The damage evolution is visible in Table~\ref{tb:Platef}. Here, the expectation and standard deviation of the damage variable $f$ are presented for three different time points. As the results for the expectation are nearly indistinguishable, we only report the results of the TSM. Over time, the damage evolves mainly in the right half of the plate. Especially the region on the right side of the hole is damaging quickly.
The results of the standard deviation show a high similarity between the TSM and MC.

\begin{figure}
	\centering
	\begin{subfigure}[b]{0.32\textwidth}
		\centering
		\includegraphics[width=\textwidth]{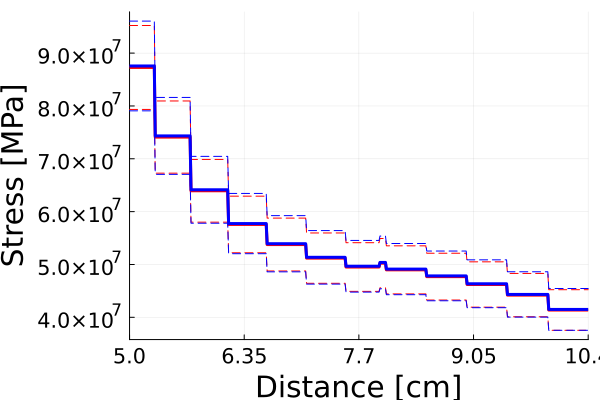}
		\caption{0.25 s}
	\end{subfigure}
	\hfill
	\begin{subfigure}[b]{0.32\textwidth}
		\centering
		\includegraphics[width=\textwidth]{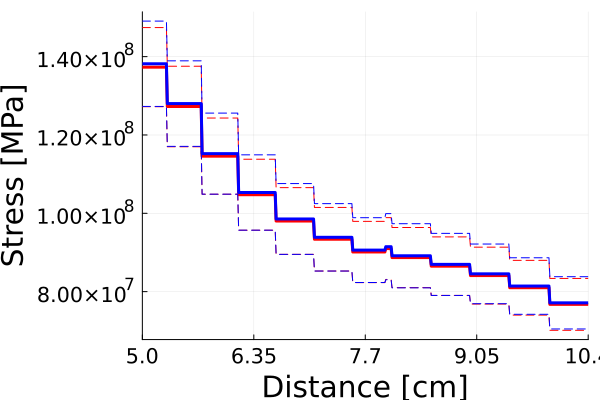}
		\caption{0.5 s}
	\end{subfigure}
	\hfill
	\begin{subfigure}[b]{0.32\textwidth}
		\centering
		\includegraphics[width=\textwidth]{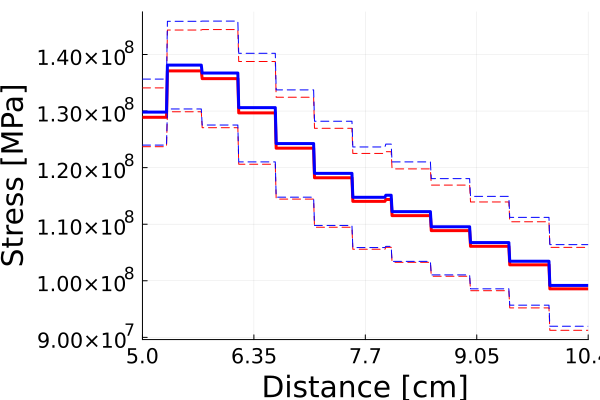}
		\caption{0.75 s}
	\end{subfigure}
	\caption{Results for the stress $\bfsigma$ along the line as indicated in Figure~\ref{fig:DNSGeometry} at three time points. The expectation $|| \langle \bfsigma \rangle ||$ is given by a solid line, the results for $|| \langle \bfsigma \rangle || \pm || \textrm{Std}(\bfsigma) ||$ by dashed lines. The results of the TSM are given in blue and the results of MC in red.}
	\label{fig:PlateStress}
\end{figure}

We additionally report the stress along a line in Figure~\ref{fig:PlateStress}. The orientation of the line is indicated in blue in Figure~\ref{fig:PlateGeometry}.
The expectation $||\langle \bfsigma \rangle ||$ is indicated by a solid line, whereas $||\langle \bfsigma \rangle|| \pm ||\textrm{Std}(\bfsigma)||$ is depicted by a dashed line. The results of the TSM are presented in blue and the results of MC in red. The average of the stress value for each cell is reported such that a step-shaped pattern results. The results are nearly identical. Interesting to observe is the decrease of the stress at $t = \SI{0.75}{s}$ at the left. The increase of the standard deviation during the damage evolution is approximated very well by the TSM.

\begin{figure}
	\centering
	\includegraphics[scale=0.4]{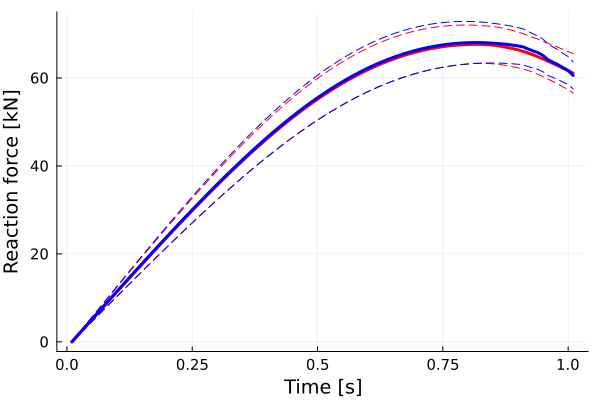}
	\caption{Reaction force in direction of the displacement over time. The expectation $\langle F_y \rangle$ is given by a solid line, the results for $\langle F_y \rangle \pm \textrm{Std}(F_y) $ by dashed lines. The results of the TSM are given in blue and the results of MC in red.}
	\label{fig:PlateReactionForce}
\end{figure}

The reaction force is an important quantity to quantify the evolution of damage in the specimen. The results for TSM and MC are presented in Figure~\ref{fig:PlateReactionForce}. The decrease in stiffness due to the damage evolution is clearly visible in the reaction force. 
The results for TSM and MC are nearly identical for expectation and standard deviation.

\begin{table}
	\caption{Evolution of the damage variable $f$ for three time points. The results for the expectation $\langle f \rangle$ are visually identical for TSM and MC. The results for the standard deviation $\textrm{Std}(f)$ are distinguished between TSM and MC.}
	\begin{tabular}{|p{2cm}|p{4cm}p{4cm}p{4cm}p{1cm}}
		\hline & 0.25s & 0.5s & 0.75s & \\
		\hline \centered{$\langle f \rangle$} & \centered{\includegraphics[scale=0.25, trim=19cm 3cm 20cm 3cm,clip]{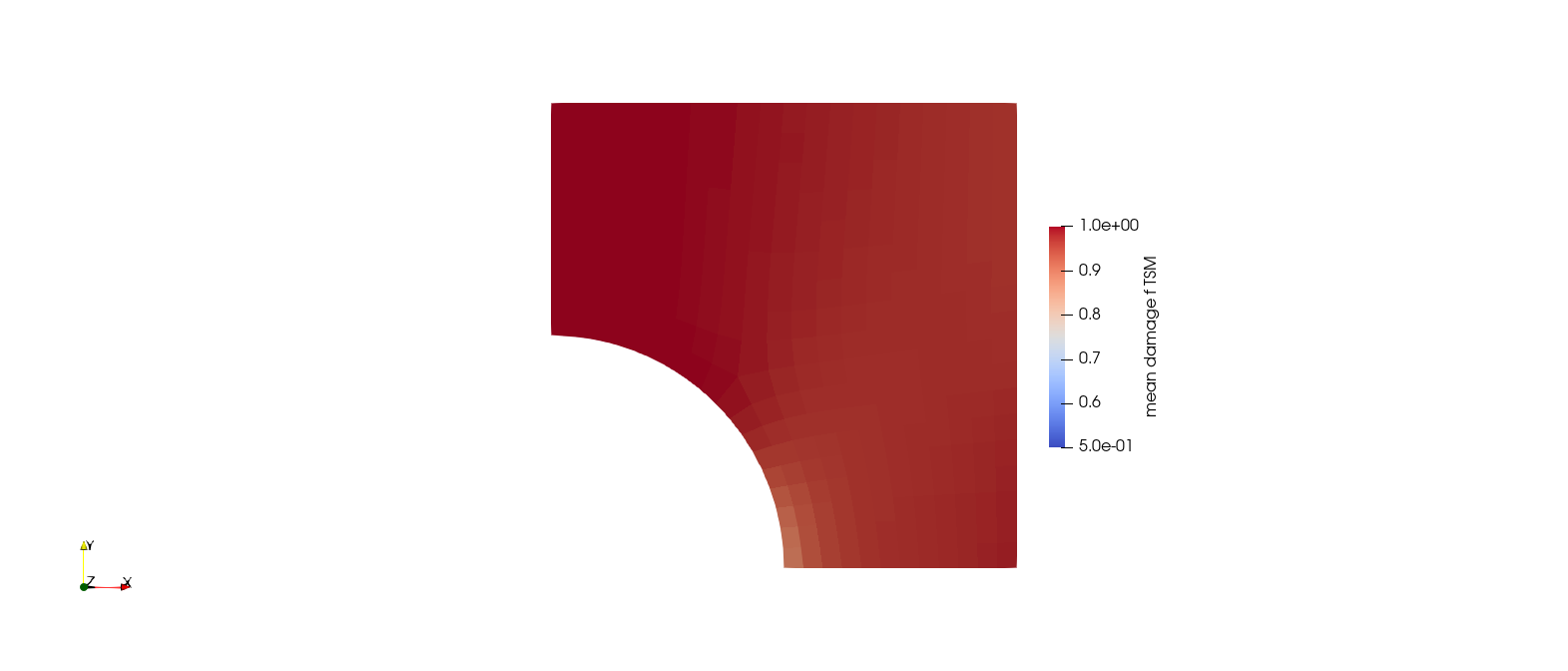}} & \centered{\includegraphics[scale=0.25, trim=19cm 3cm 20cm 3cm,clip]{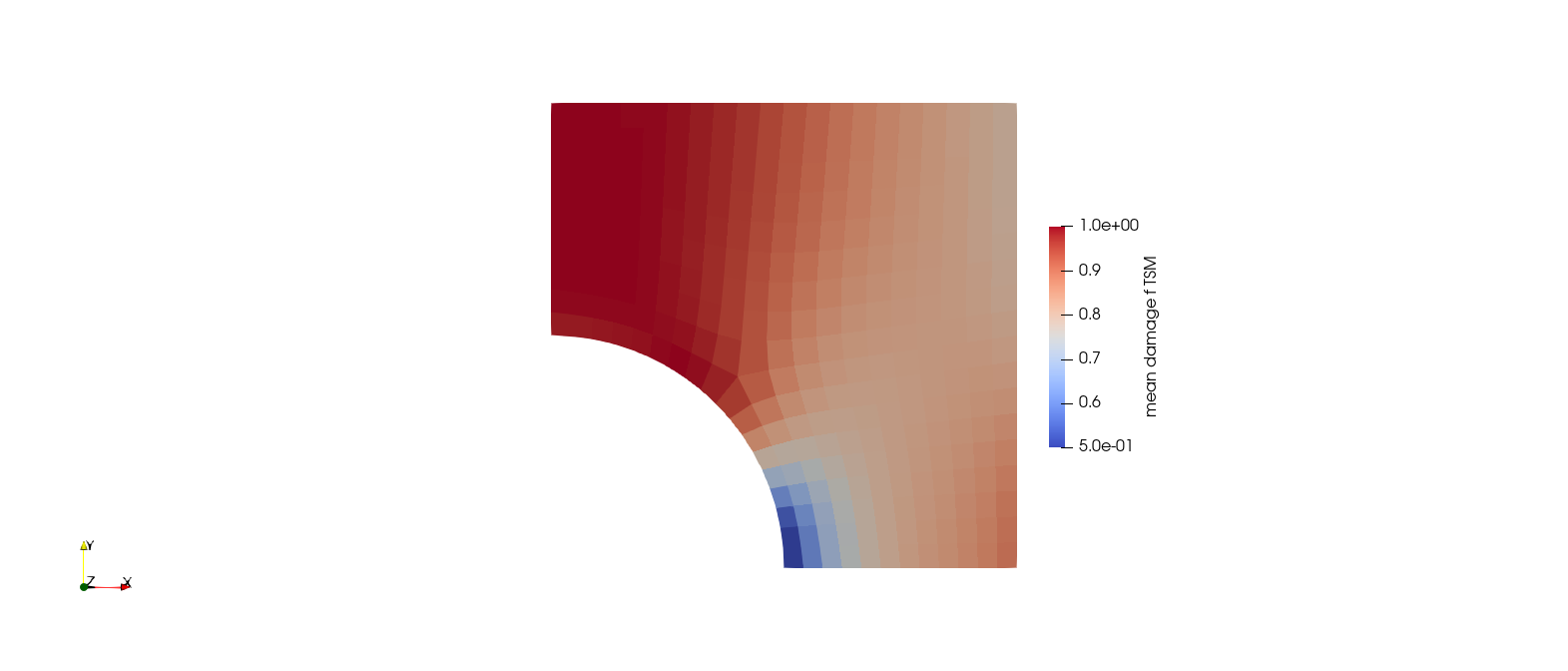}} & \centered{\includegraphics[scale=0.25, trim=19cm 3cm 20cm 3cm,clip]{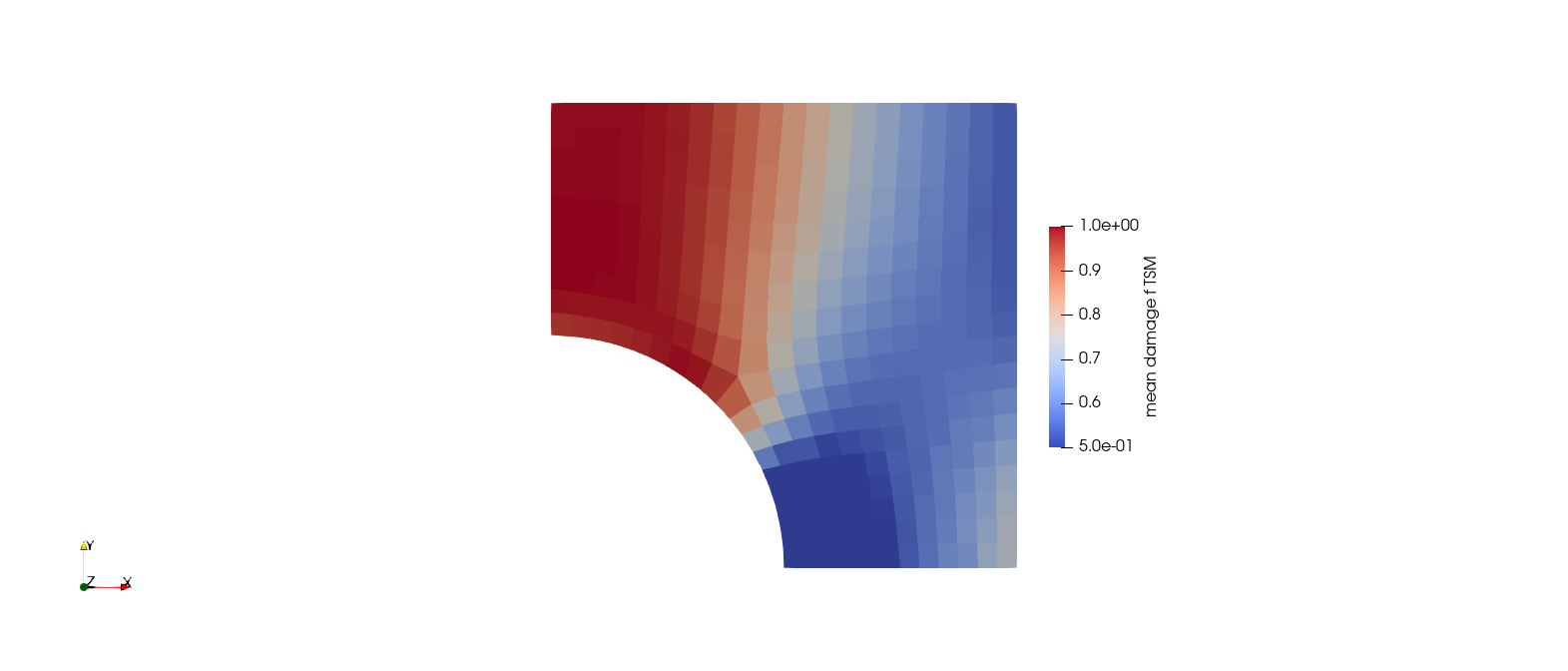}} & \centered{\includegraphics[scale=0.4, trim=36cm 7cm 15cm 7cm,clip]{plots/Round/mean_f_TSM_99.png}}\\ 
		
		\hline \centered{$\textrm{Std}(f)$ \\ by MC} & \centered{\includegraphics[scale=0.25, trim=19cm 3cm 20cm 3cm,clip]{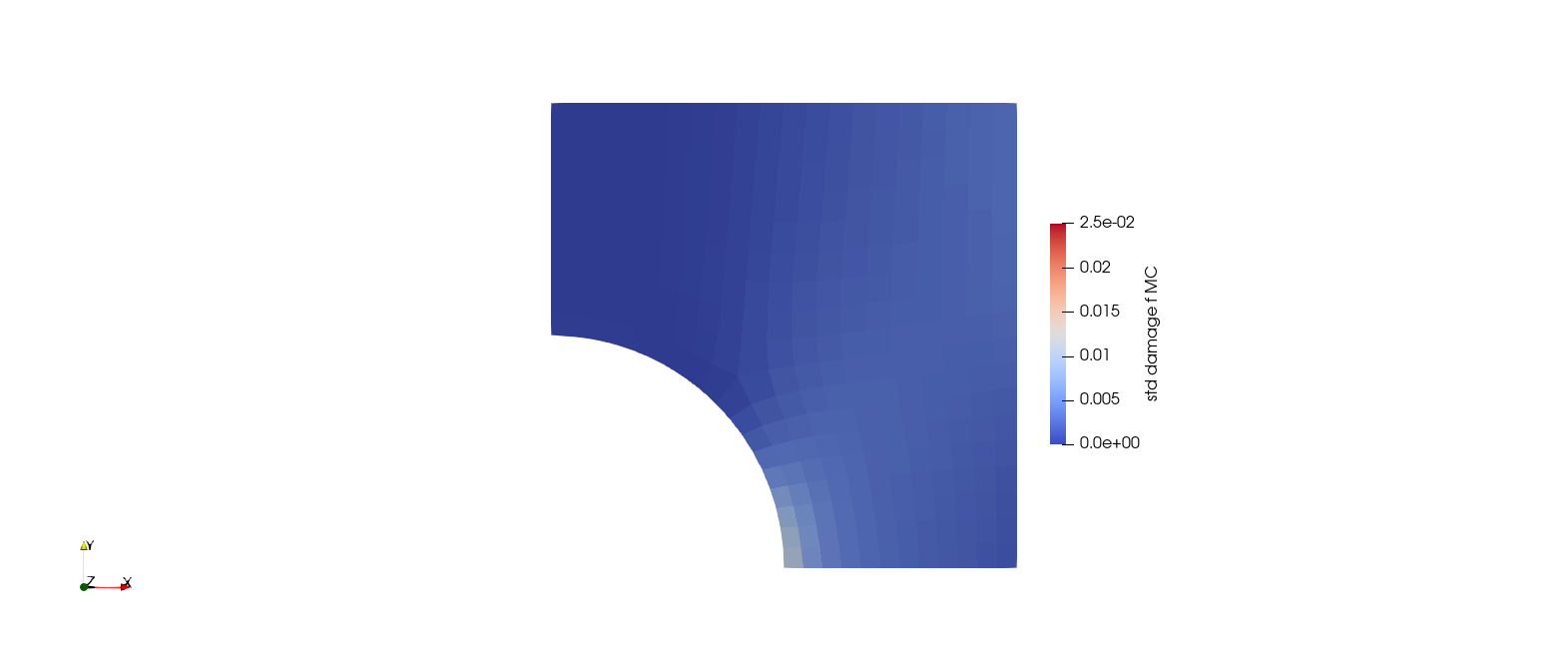}} & \centered{\includegraphics[scale=0.25, trim=19cm 3cm 20cm 3cm,clip]{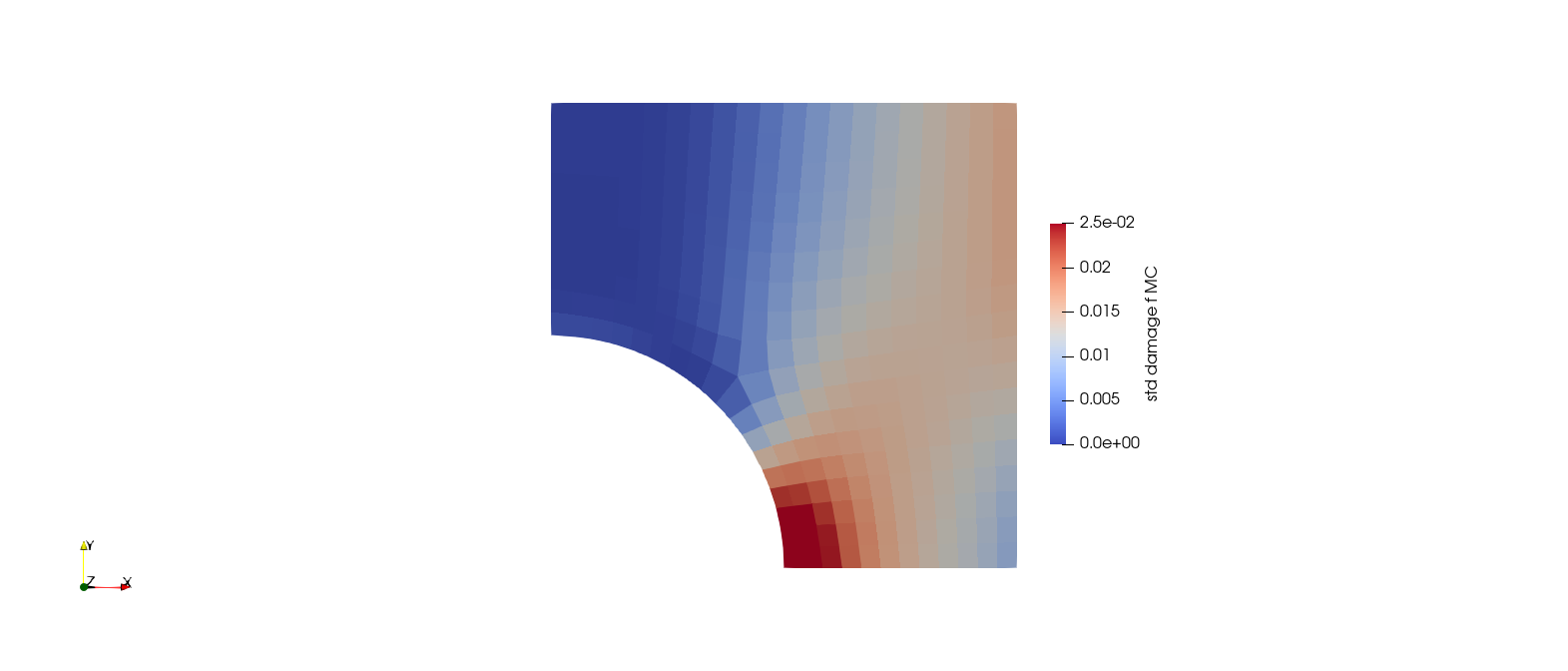}} & \centered{\includegraphics[scale=0.25, trim=19cm 3cm 20cm 3cm,clip]{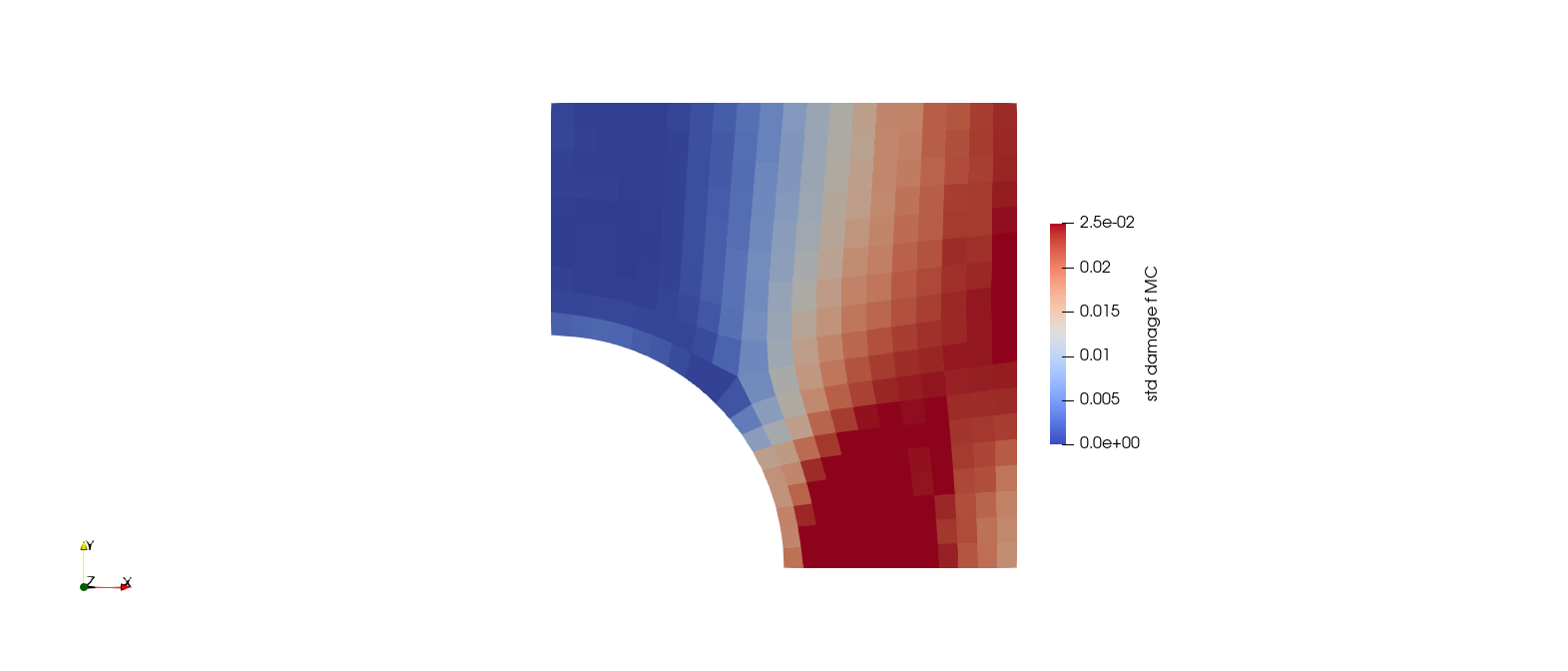}} & \centered{\includegraphics[scale=0.4, trim=36cm 7cm 15cm 7cm,clip]{plots/Round/std_f_MC_99.png}}\\
		
		\hline \centered{$\textrm{Std}(f)$ \\ by TSM} & \centered{\includegraphics[scale=0.25, trim=19cm 3cm 20cm 3cm,clip]{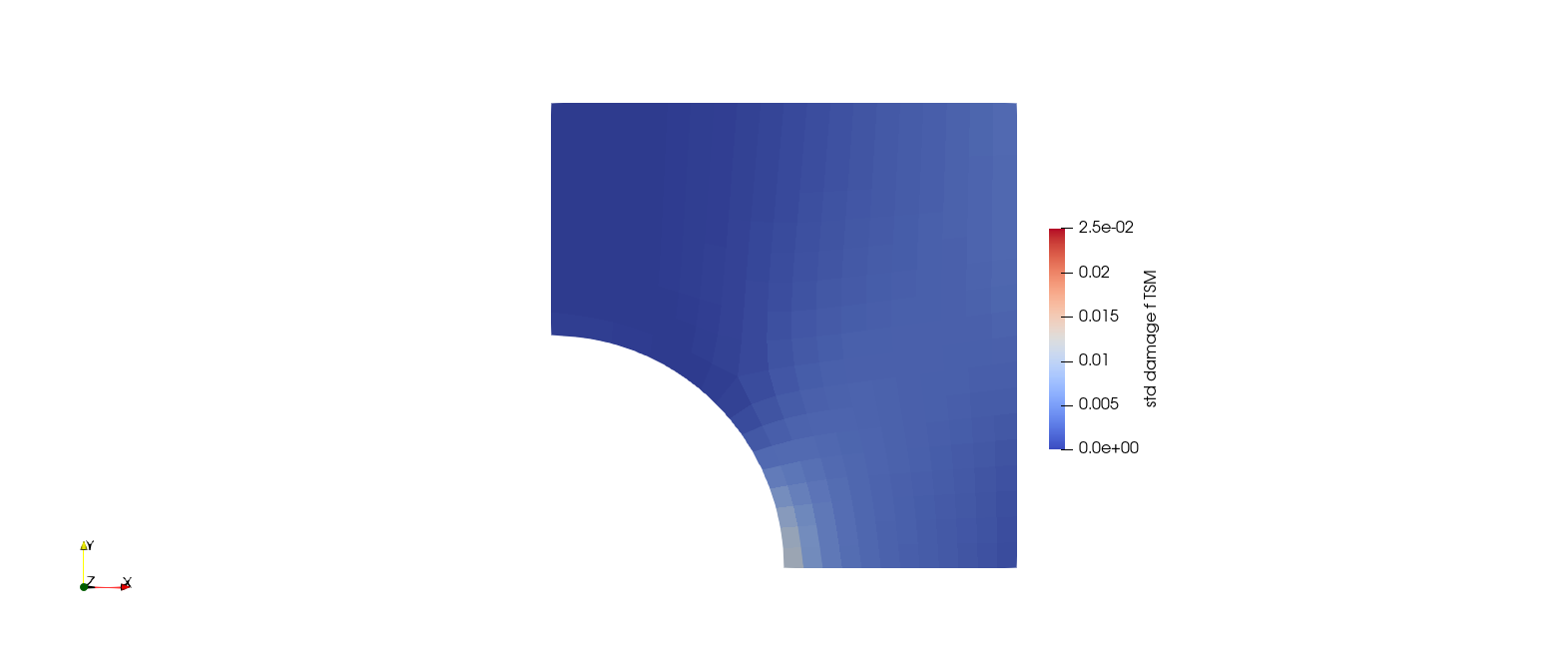}} & \centered{\includegraphics[scale=0.25, trim=19cm 3cm 20cm 3cm,clip]{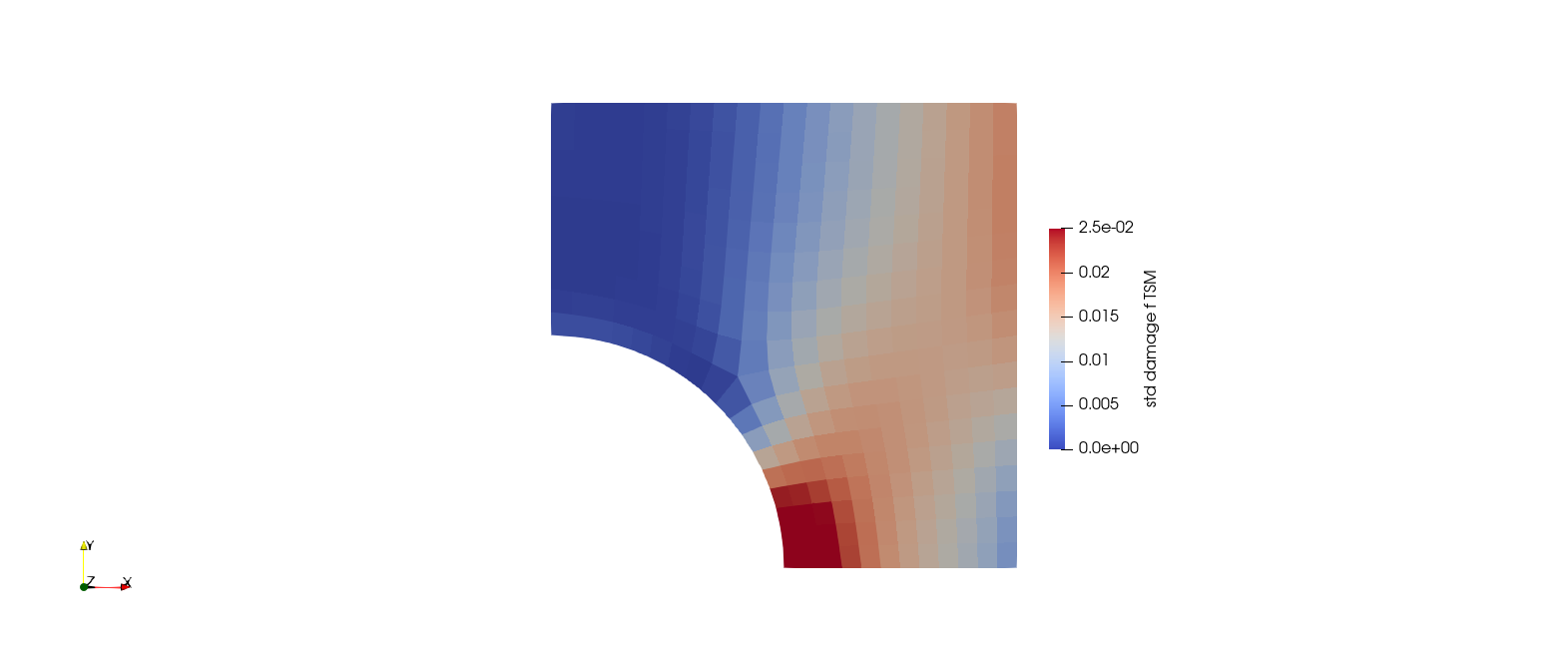}} & \centered{\includegraphics[scale=0.25, trim=19cm 3cm 20cm 3cm,clip]{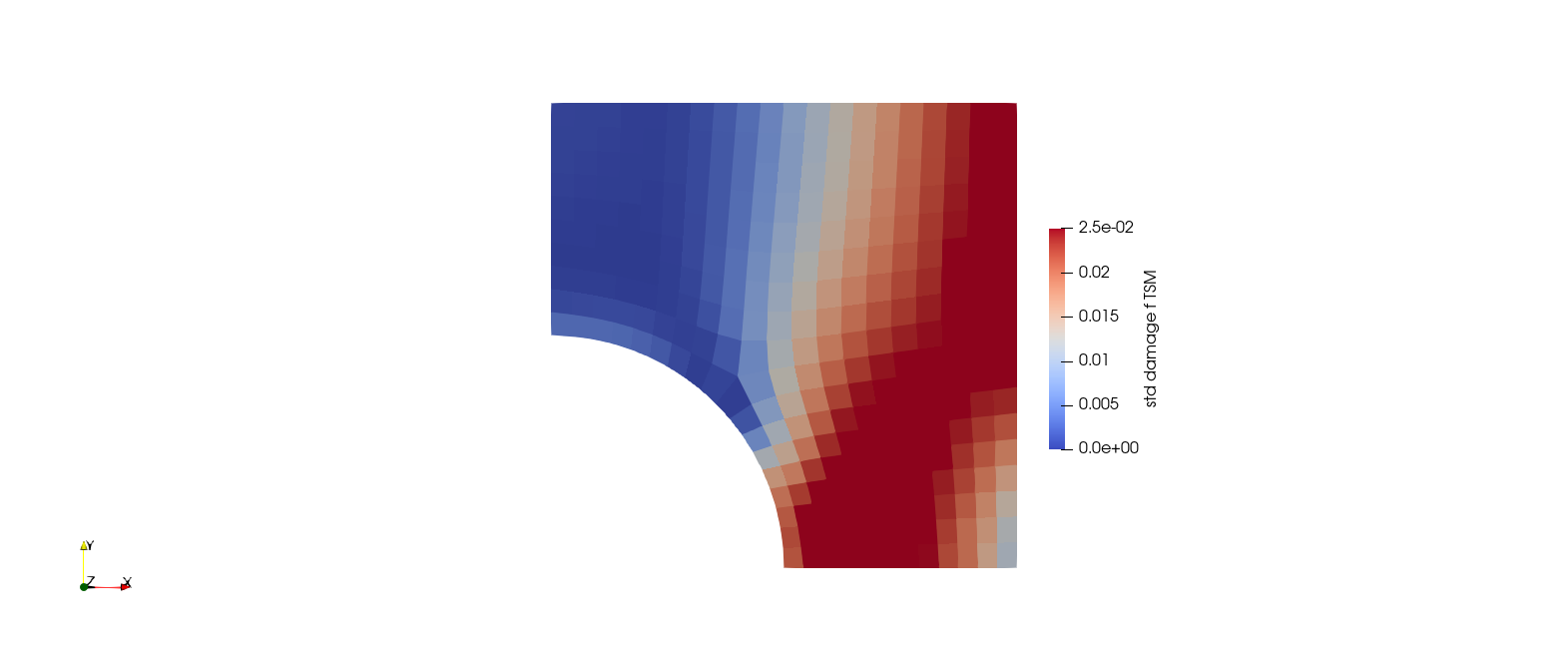}} & \centered{\includegraphics[scale=0.4, trim=36cm 7cm 15cm 7cm,clip]{plots/Round/std_f_TSM_99.png}}\\
		\hline 
	\end{tabular} 
	\label{tb:Platef}
\end{table}

\section{Computational Efficiency}
\label{sec:CompEff}
In the last Section~\ref{sec:NumericalResults}, the accuracy of the method was investigated in detail. It remains to quantify the computational efficiency of the TSM. As the number of Monte Carlo sampling points can be chosen arbitrarily as long as the results are converged, we first compare against a single deterministic calculation. The results of the comparison are presented in Table~\ref{tb:CompTimes}.
All simulations were performed on a system with a Intel Xeon Gold 6346 CPU at $\SI{3.1}{GHz}$.
Here, we report the computation times for both boundary value problems. It may be remarked, that a single deterministic simulation does not allow for any statement on the stochastic characteristics of the output fields. The TSM needs two simulations which each take $23 - 29\%$ longer 
than a single deterministic simulation. This is due to the additional file writing/reading operations as described in Section~\ref{sec:ImplementationAbaqus}. The post-processing of the results takes approximately the same amount of time as a single deterministic simulation. The main computational cost during the post-processing is the passing of results from the Abaqus Output files (\texttt{.odb}) into memory in the Python Scripting Engine.  In total, the TSM needs the same amount of time as approximately $3.5 - 3.8$ deterministic simulations. A Monte Carlo simulation with this number of sampling points would not be converged.

In fact, it is worth to compare against the Monte Carlo method in more detail. For the Monte Carlo simulation, the results of each simulation have to be passed into memory by a Python Script. This leads to an overhead, such that each individual Monte Carlo simulation is taking slightly ($\approx \SI{5}{s}$) more time than one of the TSM simulations. However, the post-processing, i.e., the calculation of expectation and standard deviation is very fast, taking less than one minute. In total, for 500 sampling points, $\SI{27}{h}$ for the double-notched specimen and $\SI{21.5}{h}$ for the plate with a hole are needed. The TSM needs less than $0.6 \%$ of this time, resulting in a speed-up factor of $170 - 180$ for the TSM. 

\begin{table}
	\centering
	\caption{Computation times of the TSM compared to a single deterministic calculation.}
	\label{tb:CompTimes}
 	\begin{tabular}{p{2cm}|c||c|c|c}
		 & Deterministic & \multicolumn{3}{c}{Time-separated stochastic mechanics} \\ 
		\hline  & Total & Simulations (each) & Post-Processing & Total \\ 
		\hline Double-notched specimen & \SI{2}{min} \SI{35}{s} & \SI{3}{min} \SI{10}{s} & \SI{2}{min} \SI{40}{s} & \SI{9}{min} \\ 
		\hline Plate with hole & \SI{2}{min} & \SI{2}{min} \SI{35}{s} & \SI{2}{min} \SI{30}{s} & \SI{7}{min} \SI{40}{s} \\ 
	\end{tabular} 	
\end{table}

\section{Conclusions}
\label{sec:Conclusion}
In this work, a \rev{weakly}-intrusive implementation of the time-separated stochastic mechanics in the finite element software Abaqus is presented for viscous damage simulations. The time-separated stochastic mechanics reduces the computational complexity of uncertainty quantification for structures with inelastic material behavior drastically. In fact, only two FEM simulations are needed to approximate the effect of a homogeneous uncertain material parameter fluctuation. In the investigated boundary value problems, the time-separated stochastic mechanics is able to approximate expectation and standard deviation of stress, damage variable and reaction force with high accuracy especially during the onset of damage. The speed-up for the investigated problems is $\approx 170$ compared to reference Monte Carlo simulations. 
The presented implementation in Abaqus allows the usage of this promising method for a larger audience and enables the implementation of uncertainty quantification also for industrial scale problems.

\section*{Acknowledgment}
This work has been supported by the German Research Foundation (DFG) 
within the framework of the international research training group IRTG 
2657 "Computational Mechanics Techniques in High Dimensions" (Reference: 
GRK 2657/1, Project number 433082294).

\appendix
%
%

\addcontentsline{toc}{chapter}{Bibliography}
\printbibliography

\end{document}

%% file: Figures/LiteratureReview.pdf_tex
\begingroup%
  \makeatletter%
  \providecommand\color[2][]{%
    \errmessage{(Inkscape) Color is used for the text in Inkscape, but the package 'color.sty' is not loaded}%
    \renewcommand\color[2][]{}%
  }%
  \providecommand\transparent[1]{%
    \errmessage{(Inkscape) Transparency is used (non-zero) for the text in Inkscape, but the package 'transparent.sty' is not loaded}%
    \renewcommand\transparent[1]{}%
  }%
  \providecommand\rotatebox[2]{#2}%
  \newcommand*\fsize{\dimexpr\f@size pt\relax}%
  \newcommand*\lineheight[1]{\fontsize{\fsize}{#1\fsize}\selectfont}%
  \ifx\svgwidth\undefined%
    \setlength{\unitlength}{451.77128129bp}%
    \ifx\svgscale\undefined%
      \relax%
    \else%
      \setlength{\unitlength}{\unitlength * \real{\svgscale}}%
    \fi%
  \else%
    \setlength{\unitlength}{\svgwidth}%
  \fi%
  \global\let\svgwidth\undefined%
  \global\let\svgscale\undefined%
  \makeatother%
  \begin{picture}(1,0.53540978)%
    \lineheight{1}%
    \setlength\tabcolsep{0pt}%
    \put(0,0){\includegraphics[width=\unitlength,page=1]{LiteratureReview.pdf}}%
    \put(0.7672183,0.07193033){\color[rgb]{0,0,0}\makebox(0,0)[lt]{\lineheight{1.25}\smash{\begin{tabular}[t]{l}accuracy of\\material model\end{tabular}}}}%
    \put(0.46602277,0.00552512){\makebox(0,0)[lt]{\lineheight{0}\smash{\begin{tabular}[t]{l}physically correct\end{tabular}}}}%
    \put(0.19684425,0.00552512){\makebox(0,0)[lt]{\lineheight{0}\smash{\begin{tabular}[t]{l}simplified\end{tabular}}}}%
    \put(-0,0.5118653){\makebox(0,0)[lt]{\lineheight{0}\smash{\begin{tabular}[t]{l}computation speed\\ \\ \\\\\end{tabular}}}}%
    \put(0.0450607,0.39209862){\makebox(0,0)[lt]{\lineheight{0}\smash{\begin{tabular}[t]{l}high\end{tabular}}}}%
    \put(0.04496991,0.12173429){\makebox(0,0)[lt]{\lineheight{0}\smash{\begin{tabular}[t]{l}low\end{tabular}}}}%
    \put(0,0){\includegraphics[width=\unitlength,page=2]{LiteratureReview.pdf}}%
    \put(0.18972936,0.38808592){\makebox(0,0)[lt]{\lineheight{0}\smash{\begin{tabular}[t]{l}Perturbation\end{tabular}}}}%
    \put(0,0){\includegraphics[width=\unitlength,page=3]{LiteratureReview.pdf}}%
    \put(0.2879408,0.24980157){\makebox(0,0)[lt]{\lineheight{0}\smash{\begin{tabular}[t]{l}Stochastic Collocation\end{tabular}}}}%
    \put(0,0){\includegraphics[width=\unitlength,page=4]{LiteratureReview.pdf}}%
    \put(0.30691375,0.11580526){\makebox(0,0)[lt]{\lineheight{0}\smash{\begin{tabular}[t]{l}Sampling Method\end{tabular}}}}%
    \put(0,0){\includegraphics[width=\unitlength,page=5]{LiteratureReview.pdf}}%
    \put(0.43756332,0.40039924){\makebox(0,0)[lt]{\smash{\begin{tabular}[t]{l}Time-separated\\stochastic mechanics\\\end{tabular}}}}%
  \end{picture}%
\endgroup%